\documentclass[10pt,journal,doublecolumn]{IEEEtran}
\usepackage{threeparttable,placeins,makecell,stfloats,cite}
\usepackage{threeparttable,arydshln}
\usepackage{amsmath,amssymb}
\usepackage[dvips]{graphicx}
\usepackage[usenames]{color}
\usepackage{amsfonts}
\usepackage{latexsym}
\usepackage{multirow}
\usepackage{caption}
\usepackage{rotating}
\usepackage{float}
\usepackage[format=hang]{subfig}
\newtheorem{lemma}{{\textit{{Lemma}}}}

\newtheorem{definition}{\textit{{Definition}}}

\newtheorem{remark}{\textit{{Remark}}}
\newtheorem{proof}{\textit{{Proof}}}

\newtheorem{example}{\textit{{Example}}}
\newtheorem{construction}{\textit{{Construction}}}

\begin{document}

\title{Cross Z-Complementary Pairs (CZCPs) for Optimal Training in Spatial Modulation Over Frequency Selective Channels}
\author{Zilong~Liu,~Ping Yang,~Yong Liang Guan,~Pei Xiao
\thanks{Zilong Liu is with School of Computer Science and Electronic Engineering, University of Essex, United Kingdom (e-mail: {\tt zilong.liu@essex.ac.uk}). Ping Yang is with National Key Laboratory on Communications, University of Electronic Science and Technology of China, Chengdu, China (e-mail: {\tt yang.ping@uestc.edu.cn}). Pei Xiao is with Institute for Communication Systems, 5G Innovation Centre, University of Surrey, United Kingdom (e-mail: {\tt p.xiao@surrey.ac.uk}). Yong Liang Guan is with the School of Electrical and Electronic Engineering, Nanyang Technological University, Singapore (E-mail: {\tt eylguan@ntu.edu.sg}).}
}

\maketitle

\begin{abstract}
The contributions of this paper are twofold: Firstly, we introduce a novel class of sequence pairs, called ``cross Z-complementary pairs (CZCPs)", each displaying zero-correlation zone (ZCZ) properties for \textit{both} their aperiodic autocorrelation sums and crosscorrelation sums. Systematic constructions of {\textit{perfect}} CZCPs based on selected Golay complementary pairs (GCPs) are presented. Secondly, we point out that CZCPs can be utilized as a key component in designing training sequences for \textit{broadband} spatial modulation (SM) systems. We show that our proposed SM training sequences derived from CZCPs lead to \textit{optimal} channel estimation performance over frequency-selective channels.
\end{abstract}


\section{Introduction}
This paper investigates a novel class of sequence pairs called ``cross Z-complementary pairs (CZCPs)" and their applications for training sequence design in \textit{broadband} spatial modulation (SM) systems. In what follows, we review the state-of-the-art on pairs of sequences and SM, then introduce our contributions in this work.

\subsection{Pairs of Sequences}
``Pairs of sequences" has been an attractive research topic since the invention of complementary pairs proposed by Marcel J. E. Golay in 1951 \cite{Golay-51}. It was found by Golay that infrared multislit spectrometry, a device which isolates a desired radiation with a fixed single wavelength from background radiation (with many different wavelengths), can be designed with the aid of a special class of pairs of sequences, which is widely known as Golay complementary pair (GCP).

Let $\mathbf{a}$ and $\mathbf{b}$ be two sequences of identical length. Denote by $\rho(\mathbf{a},\mathbf{b})(\tau)$ their aperiodic crosscorrelation (ACC) at time-shift $\tau$ (formal definition of ACC will be given in Section II). For simplicity, when $\mathbf{a}=\mathbf{b}$, $\rho(\mathbf{a},\mathbf{b})(\tau)$ will be written as $\rho(\mathbf{a})(\tau)$. By definition, $(\mathbf{a},\mathbf{b})$ is said to be a GCP if $\rho(\mathbf{a})(\tau)+\rho(\mathbf{b})(\tau)=0$ for all $\tau\neq0$. In principle, sequences in a GCP work in a cooperative way so that their aperiodic autocorrelations (AAC) sums vanish for all the non-zero time-shifts \cite{Golay61}.

For more than 60 years, huge amount of research effort has been made on the constructions of GCPs. Recursive algorithms proposed by Budi\v{s}in in 1990s for polyphase and multi-level GCPs can be found in \cite{Budisin90,Budisin902}. Theory on para-unitary (PU) matrices has recently been applied for more efficient GCP synthesis \cite{Budisin14,Wang-SETA16,Das18,Budisin18}. In 1999, Davis and Jedwab developed a direct construction of polyphase GCPs through generalized Boolean functions \cite{Davis-Jedwab1999}. Their construction was extended to GCPs over QAM constellations using weighted sum of several QPSK GCPs \cite{Li-GCS-2013,Liu-GCS-2013}. Existing known binary GCPs have even lengths of the form $2^\alpha10^\beta26^\gamma$ only, where $\alpha,\beta,\gamma$ are non-negative integers satisfying $\alpha+\beta+\gamma\geq 1$ \cite{Fan-book}, \cite{Parker02}. For sequence pairs with odd lengths, the best known alternative to GCPs is \textit{optimal} odd-length binary Z-complementary pairs (OB-ZCPs) \cite{Liu-2014} each displaying the largest zero-correlation zone (ZCZ) width \cite{Fan07} and minimum AAC sums outside the ZCZ.

GCPs and their extension, called ``complementary codes (CC)" \cite{Tseng72,Rath08}, have found applications far beyond their initial usage in the design of multislit spectrometry. A few selected applications in wireless communications include: optimal channel estimation in multiple-input multiple-output (MIMO) frequency-selective channels \cite{SQWang2005,SQWang2007}, Doppler-resilient waveform design \cite{Budisin91,AAWS2008}, interference-free multi-carrier code-division multiple-access (MC-CDMA) communications \cite{HHC01,HHC-book,Liu-TWC2014}, and power control in MC communications \cite{Davis-Jedwab1999,Li-GCS-2013,Liu-GCS-2013,Popovic91,Paterson00,Liu-TCOM2013}.

It should be noted that each GCP, ZCP or CC, is defined by their AAC sums only. Within each GCP, ZCP or CC, it is assumed that separate non-interfering channels are used for the transmission of the constituent sequences. Crosscorrelations between constituent sequences within each GCP, ZCP or CC, have not been considered.
\subsection{Spatial Modulation (SM)}
 MIMO is a revolutionary communication paradigm which can achieve significant increase in spectral efficiency and enhanced robustness against fading through exploitation of spatial diversity \cite{Telatar99,Tarokh99,Hassibi02,Zheng03,Mietzner09,Basar16,Zheng17}. SM is a special class of MIMO techniques which trades multiplexing gain with complexity and performance \cite{Mesleh05,Mesleh08,Renzo11,Yang15,Wen2019}. Unlike conventional MIMO, an SM system is equipped with multiple transmit antenna (TA) elements but only a single radio-frequency (RF) chain. During each time-slot, an SM symbol can be divided into two parts: one (called ``spatial symbol") is responsible for the selection and then activation of a TA element, the other (called ``constellation symbol") is selected from a conventional PSK/QAM constellation and transmitted from the active TA element. Such unique transmission principle of SM allows it to have at least the following advantages over conventional MIMO:
\begin{enumerate}
\item Zero inter-channel interference (ICI) over flat fading channels and thus improved average bit error probability;
\item Lower hardware complexity in the transmitter (due to its single RF chain) and lower detection complexity at the receiver (due to zero ICI);
\item Lower energy consumption which makes SM a promising candidate for large-scale antenna systems \cite{Yang16}.
\end{enumerate}

Early literature on SM mostly assume that channel state information (CSI) is perfectly known at the receiver \cite{Jeganathan08,Renzo12}. Effects of channel estimation errors to the performance of SM systems, limited to flat-fading channels, have been studied in \cite{Basar12,Renzo12v2,Sugiura12,Wu2014}. Up to date, however, little has been understood on channel estimation of SM in frequency-selective channels. Along this research direction, two fundamental research problems arise and will be addressed in this work: 1) What is the lower bound on the channel estimation mean-squares-error (MSE)? 2) How to design efficient channel training scheme to meet this lower bound with equality?

Note that the ``one-RF-chain" principle of SM prevents the transmitter from using simultaneous pilot transmission over all the TAs. Consequently, it implies that dense training sequences proposed in \cite{Yang02,Fragouli03,Fan04} for traditional MIMO are unapplicable in SM systems. Although an identity training matrix has been employed for joint channel estimation and data detection in SM systems \cite{Sugiura12}, extension to frequency-selective channels is not straightforward. A naive scheme is to extend a perfect sequence (having zero autocorrelation sidelobes) with cyclic-prefix (CP)\footnote{whose duration should be not smaller than the largest multipath delay.} and then send the extended sequence in turn over multiple TAs. But this training scheme would be inefficient in highly dispersive channels.

\subsection{Contributions of This Paper}
The main contributions of this work are summarized as follows.
\begin{enumerate}
\item We introduce a novel class of sequence pairs, called \textit{CZCPs}, each displaying certain ZCZ properties for both their AAC sums and ACC sums. Specifically, a pair of sequences, say $(\mathbf{a},\mathbf{b})$, is called a CZCP if $\rho(\mathbf{a})(\tau)+\rho(\mathbf{b})(\tau)$ and $\rho(\mathbf{a},\mathbf{b})(\tau)+\rho(\mathbf{b},\mathbf{a})(\tau)$ take zero values for certain time-shifts $\tau$. The most distinctive feature of CZCPs (compared to conventional sequence pairs) is that the two constituent sequences in a CZCP may interfere with each other during the transmission. Therefore, careful design should be carried out to force their cross-interference to zero in a certain $\tau$ region. We investigate the structural properties of CZCPs and show that a subset of GCPs are  {\textit{perfect}} CZCPs.
\item We present a generic training framework for SM training over frequency selective channels. Under this framework, we derive the lower bound on channel estimation MSE using least square (LS) estimator and conditions to meet this lower bound with equality. Then, we show that CZCPs play an instrumental role in the design of optimal SM training sequences (which are equivalent to certain sparse matrices). Numerical simulations indicate that the proposed SM training sequences lead to minimum channel estimation MSE w.r.t. the aforementioned lower bound.
\end{enumerate}

\subsection{Organization of This Paper}
This paper is organized as follows. Section II introduces some notations and GCPs, followed by a sketch of the basic principle of SM. In Section III, we formally define CZCPs, give their structural properties, and then present {\textit{perfect}} CZCPs obtained from computer search and systematic constructions. Section IV starts from a generic framework on the training design of SM, under which a lower bound on the channel estimation MSE is derived. Then, CZCPs are applied to design sparse training matrices for optimal channel estimation performance (w.r.t. the derived lower bound) in SM-MIMO frequency selective channels. In the end of Section IV, we evaluate the proposed SM training scheme using numerical simulations. Finally, Section V concludes this paper.

\section{PRELIMINARIES}

\subsection{Notations}
The following notations will be used throughout this paper.

\begin{itemize}
\item [$-$] $\mathbf{X}^{\text{T}}$ and $\mathbf{X}^{\text{H}}$ denote the transpose and the Hermitian transpose of matrix $\mathbf{X}$, respectively;
\item [$-$] $\|\mathbf{X}\|=\sqrt{\sum_{m=1}^{M}\sum_{n=1}^{N}|x_{m,n}|^2}$ denotes the Frobenius norm of matrix $\mathbf{X}=[x_{m,n}]_{m,n=1}^{M,N}$;
\item [$-$] $<\mathbf{x},\mathbf{y}>$ denotes the inner-product between two complex-valued sequences $\mathbf{x}=[x_0,x_1,\cdots,x_{N-1}]^{\text{T}}$ and $\mathbf{y}=[y_0,y_1,\cdots,y_{N-1}]^{\text{T}}$, i.e.,
$<\mathbf{x},\mathbf{y}>=\sum\limits_{i=0}^{N-1}x_iy^*_i$, where $N$ is the sequence length of $\mathbf{x}$ (and $\mathbf{y}$);
\item [$-$] $T^\tau({\mathbf{x}})$ denotes the right-cyclic-shift of $\mathbf{x}=[x_0,x_1,\cdots,x_{N-1}]^{\text{T}}$ for $\tau$ (nonnegative integer) positions, i.e.,
\begin{displaymath}
T^\tau({\mathbf{x}})=[\underbrace{x_{N-\tau},\cdots,x_{N-1}}_{\text{the last~}\tau~\text{elements of}~\mathbf{x}},x_0,x_1,\cdots,x_{N-\tau-1}]^{\text{T}}.
\end{displaymath}
Similarly,
\begin{displaymath}
T^{-\tau}({\mathbf{x}})=[x_\tau,x_{\tau+1},\cdots,x_{N-1},\underbrace{x_0,x_1,\cdots,x_{\tau-1}}_{\text{the first~}\tau~\text{elements of~}\mathbf{x}}]^{\text{T}}.
\end{displaymath}
\item [$-$] $\underline{\mathbf{x}}$ denotes the reversal of vector $\mathbf{x}=[x_0,x_1,\cdots,x_{N-1}]^{\text{T}}$, i.e., $\underline{\mathbf{x}}=[x_{N-1},x_{N-2},\cdots,x_0]^{\text{T}}$;
\item [$-$] $\text{Diag}[\mathbf{x}]$ denotes a diagonal matrix with the diagonal vector of $\mathbf{x}$ and zero for all off-diagonal entries;
\item [$-$] $\mathbf{1}_{m\times n}$ and $\mathbf{0}_{m\times n}$ denote an all-1 matrix and an all-0 matrix, respectively, both having matrix order of $m\times n$;
\item [$-$] $\mathbf{I}_N$ denotes the identity matrix of order $N$;
\item [$-$] $\otimes$ denotes the Kronecker product operator;
\item [$-$] $\lfloor j \rfloor_J$ denotes the modulo $J$ operation of integer $j$, e.g., $\lfloor 4 \rfloor_4=0, \lfloor 5 \rfloor_4=1$;
\item [$-$] {Denote $\omega_q=\exp\left (\frac{2\pi\sqrt{-1}}{q} \right )$ (integer $q$);}
\item [$-$] $\mathbb{Z}_q$ denotes the set of integers modulo $q$, i.e., $\mathbb{Z}_q=\{0,1,\cdots,q-1\}$ ($q$ integer);
\item [$-$] {$\mathcal{A}_q=\{\omega^0_q,\omega^1_q,\cdots,\omega^{q-1}_q\}$ denotes the set over $q$ complex roots of unity.}
\item [$-$] $\mathbb{E}(x)$ denotes the mean of random variable $x$;
\item [$-$] $\text{Tr}(\mathbf{X})$ denotes the trace of square matrix $\mathbf{X}$.
\end{itemize}

For two length-$N$ complex-valued sequences
\begin{displaymath}
\mathbf{a}=[a_0,a_1,\cdots,a_{N-1}]^{\text{T}},~~\mathbf{b}=[b_0,b_1,\cdots,b_{N-1}]^{\text{T}},
\end{displaymath}
denote by $\rho({\mathbf{a}},{\mathbf{b}})\left(\tau\right)$ the ACC between $\mathbf{a}$ and $\mathbf{b}$, i.e.,
{
\begin{equation}
\rho({\mathbf{a}},{\mathbf{b}})\left(\tau\right) =
\begin{cases}
{\sum\limits_{n = 0}^{N - \tau - 1} {a_n {b^*_{n + \tau}} }}, &~ 0 \leq \tau \leq N-1,\\
{\sum\limits_{n = 0}^{N + \tau - 1} {a_{n-\tau} {b^*_{n }} }},&~ -(N-1)\leq \tau\leq -1,\\
0, &~|\tau|\geq N.
\end{cases}
\end{equation}
}
Also, denote by $\phi({{\mathbf{a}},{\mathbf{b}}})\left(\tau\right)$ the \textit{periodic} cross-correlation (PCC) between $\mathbf{a}$ and $\mathbf{b}$, i.e.,
\begin{equation}\label{equ_periodicx}
\phi({{\mathbf{a}},{\mathbf{b}}})\left(\tau\right) = \sum\limits_{n = 0}^{N - 1} {a_n} {b^*_{\lfloor n + \tau \rfloor_{N}} }=\Bigl <\mathbf{a},T^{-\tau} (\mathbf{b}) \Bigl >.
\end{equation}
Clearly, $\rho({{\mathbf{a}},{\mathbf{b}}})\left(\tau\right)=\rho^*({{\mathbf{a}},{\mathbf{b}}})\left(-\tau\right)$ and
\begin{equation}
\phi({{\mathbf{a}},{\mathbf{b}}})\left(\tau\right)=\rho({{\mathbf{a}},{\mathbf{b}}})\left(\tau\right)+\rho^*({{\mathbf{b}},{\mathbf{a}}})\left(N-\tau\right).
\end{equation}
In particular, when {$\mathbf{a}=\mathbf{b}$}, $\rho({{\mathbf{a}},{\mathbf{a}}})(\tau)$ will be sometimes written as $\rho({{\mathbf{a}}})(\tau)$ and called the AAC of $\mathbf{a}$ at time-shift $\tau$. Similarly,  $\phi({{\mathbf{a}},{\mathbf{a}}}) (\tau)$ will be sometimes written as $\phi({{\mathbf{a}}})(\tau)$ and called the periodic auto-correlation (PAC) of $\mathbf{a}$.

\subsection{Introduction to Golay Complementary Pair (GCP)}
Next, we give a brief introduction to GCPs. We will show in Section III that a subset of GCPs are \textit{perfect} CZCPs.

\begin{definition}
$(\mathbf{a},\mathbf{b})$ is called a GCP if their AAC sum equals zero for any non-zero time-shift $\tau$, i.e.,
\begin{equation}
\rho({{\mathbf{a}})\left(\tau\right)+\rho({\mathbf{b}}})\left(\tau\right)=0,~\forall~\tau\neq 0.
\end{equation}
\end{definition}
Furthermore, two GCPs $(\mathbf{a},\mathbf{b})$ and $(\mathbf{c},\mathbf{d})$ are said to be \textit{mutually orthogonal} if they have zero ACC sums for all time-shifts, i.e.,
\begin{equation}
\rho(\mathbf{a},\mathbf{c})\left(\tau\right)+\rho(\mathbf{b},\mathbf{d})\left(\tau\right)=0,
\end{equation}
holds for any arbitrary $\tau$. As a matter of fact, $(\underline{\mathbf{b}^*},-\underline{\mathbf{a}^*})$ is a GCP which is mutually orthogonal to $(\mathbf{a},\mathbf{b})$ \cite{Fan-book}, where
$\underline{\mathbf{b}}$ denotes the reversal of $\mathbf{b}$.

{
\begin{definition}
For $\mathbf{x}=[x_1,x_2,\ldots,x_{\mu}]\in \mathbb{Z}_2^{\mu}$, a generalized Boolean function (GBF) $g(\mathbf{x})$
is defined as a mapping $g: \{0,1\}^{\mu} \rightarrow \mathbb{Z}_q$. Each variable $x_i$ ($i\in \{1,2,\ldots,\mu \}$) in $\mathbf{x}$ may be regarded as a GBF (see \textit{Example 1}). Let
$[\kappa_1,\kappa_2,\ldots,\kappa_{\mu}]$ be the binary representation of the
(non-negative) integer $\kappa=\sum_{i=1}^{\mu} \kappa_i 2^{i-1}$, with $\kappa_{\mu}$ denoting the most significant bit and $0\leq \kappa \leq 2^\mu-1$. Given $g(\mathbf{x})$, define $g_\kappa\triangleq g(\kappa_1,\kappa_2,\ldots,\kappa_{\mu})$ and its associated sequence
\begin{equation}
\mathbf{{g}}\triangleq\Bigl [g_0, g_1, \ldots, g_{2^\mu-1} \Bigl ].
\end{equation}
Furthermore, we define a complex-valued sequence associated to GBF $g$ as follows.
\begin{equation}
\varphi_q(\mathbf{g})\triangleq\Bigl [\omega_q^{g_0}, \omega_q^{g_1}, \ldots, \omega_q^{g_{2^\mu-1}} \Bigl ].
\end{equation}
\end{definition}

It is stressed that each realization of $g_{\kappa}$ is obtained by setting $x_1=\kappa_1,x_2=\kappa_2,\ldots,x_{\mu}=\kappa_{\mu}$ into the GBF $g$, for a specific $\kappa$. Hence, if $g=x_t$, where $1\leq t \leq \mu$, the associated sequence $\mathbf{g}$ is the vector formed by $g_\kappa=\kappa_t$, when $\kappa$ ranges from $0$ to $2^\mu-1$. In this case, we denote the corresponding $\mathbf{g}$ by $\mathbf{x}_t$, i.e., $\mathbf{g}=\mathbf{x}_t$. Similarly, if $g=x_{t_1}x_{t_2}$ ($t_1\neq t_2$), we write $\mathbf{g}=\mathbf{x}_{t_1}\mathbf{x}_{t_2}$, a vector obtained by element-wise multiplication of $\mathbf{x}_{t_1}$ and $\mathbf{x}_{t_2}$. We need the following definition which will be used in Section III to identify a subset of GCPs as \textit{perfect} CZCPs.
\begin{definition}\label{defi_correGBFs}
For two GBFs $g,h$ over $\mathbb{Z}_q$, we denote by $\rho_q(g,h)(\tau)$ the aperiodic correlation of their associated complex-valued sequences, i.e.,
\begin{equation}
\rho_q(g,h)(\tau)\triangleq\rho(\varphi_q(\mathbf{g}),\varphi_q(\mathbf{h}))(\tau).
\end{equation}
\end{definition}
We present the following example to illustrate the GBFs defined above. One can find it useful in understanding the GCP construction in \textit{Lemma \ref{PSK_GDJ constr_4GCP}} below.
\begin{example}
Let $\mu=3$ and $q=4$. The associated sequences of GBFs $1,x_1,x_3,2x_1x_3+1$ are shown in (\ref{equ4example1}).
\begin{figure*}
\begin{equation}\label{equ4example1}
\begin{array}{cccccccccccc}
\left [ \begin{matrix} \kappa_1 \\ \kappa_2 \\ \kappa_3 \end{matrix}\right ] & = & &\left [ \begin{matrix} 0 \\ 0 \\ 0 \end{matrix}\right ] &  \left [ \begin{matrix} 1 \\ 0 \\ 0 \end{matrix}\right ] &  {\color{blue}\left [ \begin{matrix} 0 \\ 1 \\ 0 \end{matrix}\right ]} & \left [ \begin{matrix} 1 \\ 1 \\ 0 \end{matrix}\right ] & {\color{red}\left [ \begin{matrix} 0 \\ 0 \\ 1 \end{matrix}\right ]} &  \left [ \begin{matrix} 1 \\ 0 \\ 1 \end{matrix}\right ] &  \left [ \begin{matrix} 0 \\ 1 \\ 1 \end{matrix}\right ] & \left [ \begin{matrix} 1 \\ 1 \\ 1 \end{matrix}\right ] & \\ \hline \hline
\mathbf{1}  & = &[ &1&1&1&1&1&1&1&1&],\\
\mathbf{x}_1 & =&[&0&1&{\color{blue}0}&1&{\color{red}0}&1&0&1&],\\
\mathbf{x}_3 & =&[&0&0&{\color{blue}0}&0&{\color{red}1}&1&1&1&],\\
2\mathbf{x}_1\mathbf{x}_3+\mathbf{1} & = &[& 1&1&1&1&1&3&1&3&].
\end{array}
\end{equation}
\end{figure*}
When $\kappa=2$, for instance, its binary representation is $[0,1,0]$. Hence, both the third entries of $\mathbf{x}_1$ and $\mathbf{x}_3$ take identical zero. When $\kappa=4$, its binary representation is $[0,0,1]$ and therefore, the fifth entries of $\mathbf{x}_1$ and $\mathbf{x}_3$ are zero and one, respectively. $2\mathbf{x}_1\mathbf{x}_3+\mathbf{1}$ is obtained from two times the element-wise product between $\mathbf{x}_1$ and $\mathbf{x}_3$, followed by addition with $\mathbf{1}$. As observed from (\ref{equ4example1}), the sequence of GBF $x_1$ is in fact given by that of $\kappa_1$, when $\kappa$ ranges from 0 to 7. Similarly, the sequence of GBF $x_3$ is given by that of $\kappa_3$.
\end{example}

Next, we present in \textit{Lemma 1} the GCP construction proposed by Davis and Jedwab in \cite{Davis-Jedwab1999}. \textit{Lemma 1} will be used in \textit{Construction 2} for \textit{perfect} CZCPs having lengths of power of two.
\begin{lemma} \label{PSK_GDJ constr_4GCP}(Davis-Jedwab Construction of GCP \cite{Davis-Jedwab1999})
{Let}
\begin{equation}\label{f_4GDJ_GCP}
g(\mathbf{x})\triangleq \frac{q}{2} \sum \limits_{k=1}^{\mu-1}x_{\pi(k)} x_{\pi(k+1)} + \sum
\limits_{k=1}^{\mu} w_k x_k+w,
\end{equation}
where $\pi$ is a permutation of the set $\{1,2,\ldots,\mu\}$, and $w_k,w\in\mathbb{Z}_{q}$ ($q$ even integer).
Then, for any $w'\in \mathbb{Z}_{q}$, $\varphi_q(\mathbf{{g}})$ and $\varphi_q(\mathbf{{g}}+\frac{q}{2}{\mathbf{x}}_{\pi(1)}+w'\cdot \mathbf{1})$
form a GCP over $\mathbb{Z}_{q}$ of length $2^\mu$. 
\end{lemma}
}

\subsection{Introduction to SM}
\begin{figure*}
  \centering
  \captionsetup{justification=centering}
  \includegraphics[width=5.5in]{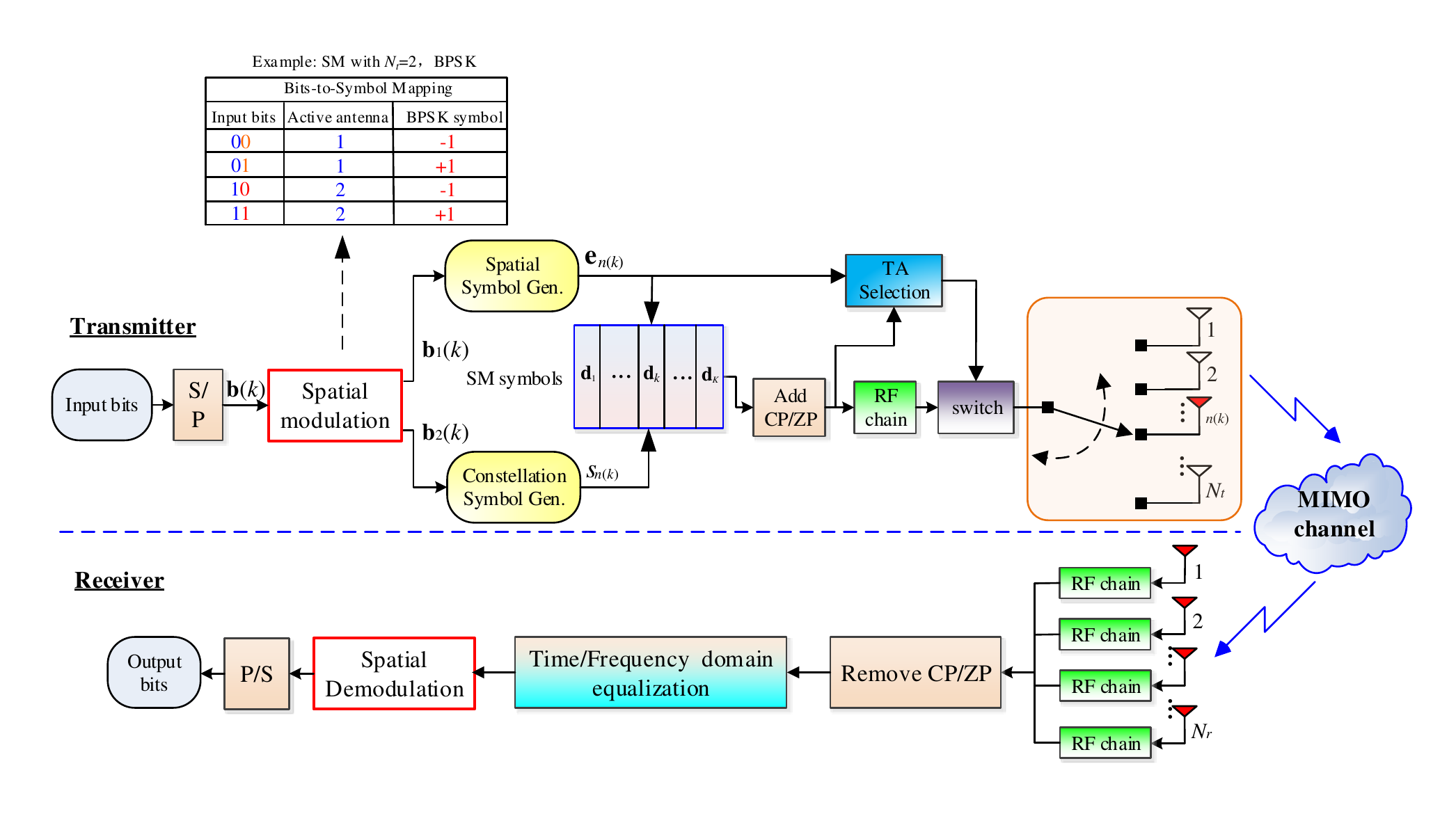}\\
  \caption{A generic transceiver structure of SC-SM systems.}
  \label{FigSC-SM}
\end{figure*}

We consider a single-carrier SM (SC-SM) system employing $N_t$ TA elements and $N_r$ receive antennas (RAs) over frequency-selective channels. Details of SC-SM and its applications in broadband large-scale antenna systems can be found in \cite{Yang16}. Moreover, we consider a QAM/PSK modulation with constellation size of $\mathcal{M}_{\text{SM}}$. Fig. \ref{FigSC-SM} shows the block diagram of a generic SC-SM transceiver. For simplicity, assume both $N_t,\mathcal{M}_{\text{SM}}$ take powers of two. Over each time-slot $k$, $\log_2(N_t\mathcal{M}_{\text{SM}})=\log_2(N_t)+\log_2(\mathcal{M}_{\text{SM}})$ bits, denoted by vector $\textbf{b}(k)$, are mapped to an SM symbol $\textbf{d}_k$. Specifically, $\log_2(N_t)$ bits, denoted by $\textbf{b}_1(k)$, are used to activate TA $n(k)$ through one RF chain. Here, we use $\textbf{e}_{n(k)}$ (which is a sparse vector corresponding to the $n(k)$-th column of the $N_t\times N_t$ identity matrix) to represent the ``spatial symbol" at time-slot $k$. On the other hand, $\log_2(\mathcal{M}_{\text{SM}})$ bits, denoted by $\textbf{b}_2(k)$, are used to select one ``constellation symbol" $S_{n(k)}$. It is noted that
\begin{equation}
\textbf{d}_k=S_{n(k)}\textbf{e}_{n(k)}.
\end{equation}

Suppose there are $K$ SM symbols (i.e., $[\textbf{d}_1,\textbf{d}_2,\cdots,\textbf{d}_K]$) in an SC block. To combat dispersive SM channels, each SC-SM block is extended by a guard interval, which may be a CP, zero prefix (ZP), or known training sequence. Signals collected from $N_r$ RAs are first processed at the receiver by removing the prefix part (in each SC-SM block). Then, time/frequency domain equalization is applied to suppress intersymbol interference (ISI). Subsequently, SM demodulation is carried out to extract spatial symbols and constellation symbols which are combined to retrieve output bits.

\begin{example}
Consider an SC-SM system with {$N_t=4$} TAs using 8-PSK modulation ($\mathcal{M}_{\text{SM}}=8$). For illustration purpose, we consider natural mapping for TA selection and 8-PSK modulation, i.e., each index (for a TA or an 8-PSK phase) can be obtained from its natural binary representation (with the left-most bit being the most significant bit). As defined in Subsection II-A, let $\omega_8=\exp(\sqrt{-1}2\pi/8)$. Suppose each SC-SM block constitutes $K=4$ SM symbols. Suppose further these symbols correspond to $K\log_2(8N_t)=20$ message bits $[10010 11011 01110 00100]$, with $\textbf{b}(1)=[10010],\textbf{b}(2)=[11011],\textbf{b}(3)=[01110]$, and $\textbf{b}(4)=[00100]$. Taking the first SM symbol for example, we have $\textbf{b}(1)=[\textbf{b}_1(1),\textbf{b}_2(1)]$, where $\textbf{b}_1(1)=[10]$ and $\textbf{b}_2(1)=[010]$. This means that during the first time-slot, only the third TA will be activated for the sending of 8-PSK symbol $\omega^2_8$. Therefore, the first SM symbol can be written as $\textbf{d}_1=[0,0,\omega^2_8,0]^{\text{T}}$. The entire SC-SM block can be expressed by the sparse matrix as follows.
\begin{equation}\label{SMexample}
[\textbf{d}_1,\textbf{d}_2,\textbf{d}_3,\textbf{d}_4]=\left [ \begin{array}{c:c:c:c}
0          & 0          & 0          & \omega^4_8\\
0          & 0          & \omega^6_8 & 0\\
\omega^2_8 & 0          & 0          & 0\\
0          & \omega^3_8 & 0          & 0\\
\end{array}
\right ].
\end{equation}
\end{example}


\section{Cross Z-Complementary Pairs: Properties and Constructions}

This section studies the properties and constructions of CZCPs. {Throughout this paper, we focus on polyphase $q$-ary CZCPs whose entries are drawn from $\mathcal{A}_q$ (see Subsection II-A for its definition).} Before proceeding further, we formally define CZCPs as follows.

\begin{definition}\label{CZCP_defi}
Let $(\mathbf{a},\mathbf{b})$ be a pair of sequences of identical length $N$. For a proper integer $Z$, define $\mathcal{T}_1\triangleq\{1,2,\cdots,Z\}$
and $\mathcal{T}_2\triangleq\{N-Z,N-Z+1,\cdots,N-1\}$. $(\mathbf{a},\mathbf{b})$ is called an $(N,Z)$-CZCP if
it possesses \textit{symmetric} zero (out-of-phase) AAC sums for time-shifts over $\mathcal{T}_1\cup \mathcal{T}_2$ and zero ACC sums for time-shifts over $\mathcal{T}_2$. In short, an $(N,Z)$-CZCP needs to satisfy the following two conditions.
\begin{equation}\label{MSE_min_ACCFcond}
\begin{split}
\text{C1}:&~\rho\left(\mathbf{a}\right)(\tau)+\rho\left(\mathbf{b}\right)(\tau)  =0,~\text{for all}~|\tau|\in \mathcal{T}_1 \cup \mathcal{T}_2;\\
\text{C2}:&~\rho\left(\mathbf{a},\mathbf{b}\right)(\tau)+\rho\left(\mathbf{b},\mathbf{a}\right)(\tau) =0,~\text{for all}~|\tau| \in \mathcal{T}_2.
\end{split}
\end{equation}
\end{definition}

From \textit{Definition \ref{CZCP_defi}}, C1 shows that each CZCP needs to have two zero autocorrelation zones (ZACZs) when its AAC sums are considered. In this paper, we call them ``front-end ZACZ" and ``tail-end ZACZ" for time-shifts over $\mathcal{T}_1$ and $\mathcal{T}_2$, respectively. On the other hand, C2 shows that each CZCP needs to have ``tail-end zero crosscorrelation zone (ZCCZ)" when its ACC sums are considered. We illustrate the correlation properties of $(N,Z)$-CZCP in Fig. \ref{CZCPex}. An example of quaternary (9,3)-CZCP is given below.
\begin{figure*}
  \centering
  \captionsetup{justification=centering}
  \includegraphics[width=6.5in]{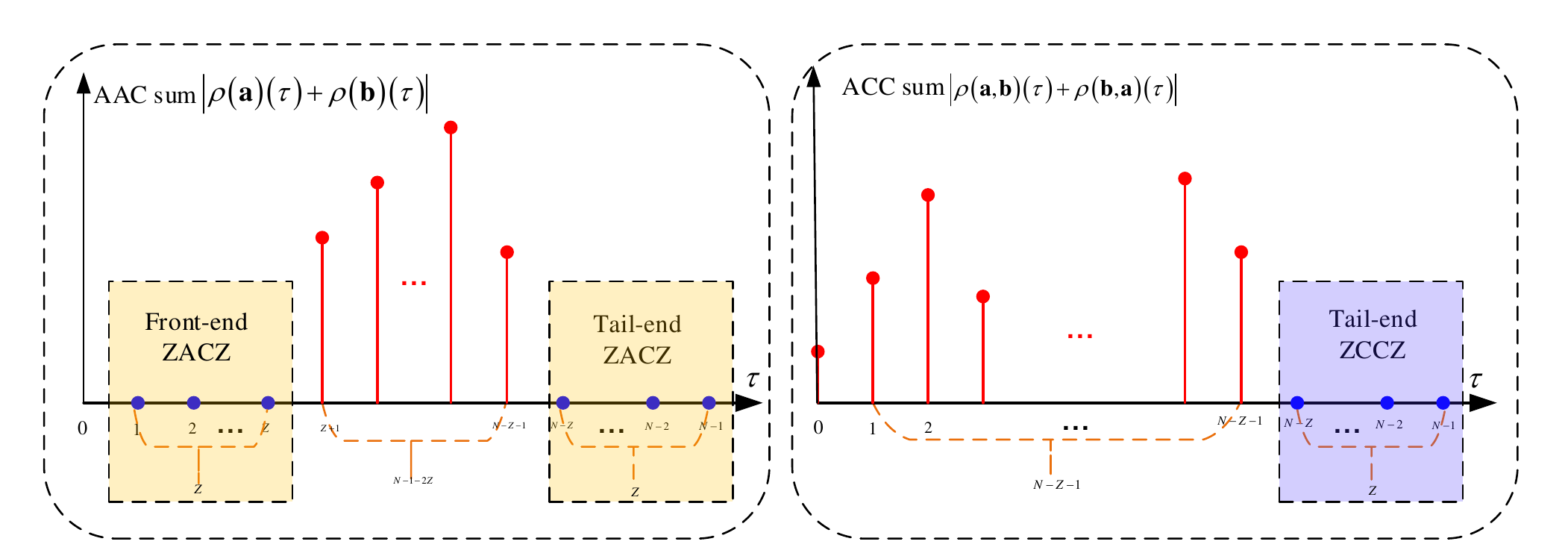}\\
  \caption{Illustrative plots for the correlation properties of $(N,Z)$-CZCP}
  \label{CZCPex}
\end{figure*}

\vspace{0.1in}
\begin{example}\label{exmp_CZCP_len9}
Consider the length-9 quaternary pair $(\mathbf{a},\mathbf{b})$ below.
{
\begin{equation}
\mathbf{a}  = \omega_4^{[0,1,1,2,0,2,1,1,3]}, ~\mathbf{b}  = \omega_4^{[0,1,1,0,1,0,3,3,1]}.
\end{equation}
}
$(\mathbf{a},\mathbf{b})$ is a $(9,3)$-CZCP because
\begin{equation}
\begin{split}
 & ~\left (\Bigl |\rho\left(\mathbf{a}\right)(\tau)+\rho\left(\mathbf{b}\right)(\tau) \Bigl | \right )_{\tau=0}^{8}\\
 = & ~(18,\mathbf{0}_{1\times3},2\sqrt{2},2,\mathbf{0}_{1\times 3}),\\
 & ~\left (\Bigl |\rho\left(\mathbf{a},\mathbf{b}\right)(\tau)+\rho\left(\mathbf{b},\mathbf{a}\right)(\tau) \Bigl | \right )_{\tau=0}^{8}\\
 = & ~\left (4,4\sqrt{2},2\sqrt{2},2\sqrt{2},4,2,\mathbf{0}_{1\times3} \right ).
\end{split}
\end{equation}

\end{example}
\vspace{0.1in}

Several main properties of CZCPs are presented below\footnote{Their proofs can be found in the Appendix of this work.}.
\vspace{0.1in}

\noindent P1: Every $q$-ary $(N,Z)$-CZCP $(\mathbf{a},\mathbf{b})$ is equivalent to $(N,Z)$-CZCP $(\mathbf{c},\mathbf{d})$
by dividing $\mathbf{a}$ by $a_0$ and $\mathbf{b}$ by $b_0$, respectively, i.e., $\mathbf{c}=\mathbf{a}/a_0, \mathbf{d}=\mathbf{b}/b_0$, where the latter CZCP satisfies
\begin{equation}\label{equ_P1}
c_i=d_i,c_{N-1-i}=-d_{N-1-i},~~\text{for all}~i\in{\{0,1,\cdots,Z-1\}}.
\end{equation}
By utilizing (\ref{equ_P1}), one can readily show that $Z\leq N/2$.

\noindent P2: If $(\mathbf{a},\mathbf{b})$ is an $(N,Z)$-CZCP, so are $(c_1{\mathbf{b}},c_2{\mathbf{a}})$, $(c_1\underline{\mathbf{b}},c_2\underline{\mathbf{a}})$, $(c_1\underline{\mathbf{b}^*},c_2\underline{\mathbf{a}^*})$, where $c_1,c_2\in \mathcal{A}_q$. Moreover, if (\ref{equ_P1}) is satisfied by $(\mathbf{a},\mathbf{b})$, i.e., $a_i=b_i,a_{N-1-i}=-b_{N-1-i},~\text{for all}~i\in{\{0,1,\cdots,Z-1\}}$, we have
\begin{equation}\label{MSE_min_ACCFcond2}
\begin{array}{cl}
\rho\left(\mathbf{a},\underline{\mathbf{b}^*}\right)(\tau)+\rho\left(\mathbf{b},-\underline{\mathbf{a}^*}\right)(\tau) & =0,~\text{for all}~\tau.\\
\rho\left(\mathbf{b},\underline{\mathbf{b}^*}\right)(\tau)+\rho\left(\mathbf{a},-\underline{\mathbf{a}^*}\right)(\tau)& =0,~\text{for all}~|\tau| \in \mathcal{T}_2.
\end{array}
\end{equation}

\noindent P3: A binary $(N,Z)$-CZCP $(\mathbf{a},\mathbf{b})$ over $\{-1,1\}$ should have even sequence length $N$ and satisfy the following equation.
\begin{equation}\label{equ_P3}
a_i+a_{N-1-i}+b_i+b_{N-1-i}=\pm 2, ~ ~ \text{for all}~i\in\{0,1,\cdots,Z-1\}.
\end{equation}



\vspace{0.1in}

\begin{definition}\label{defi_SGCP}
By P1, every $q$-ary CZCP $(\mathbf{a},\mathbf{b})$ is called  {\textit{perfect}} if $Z=N/2$ ($N$ even). In this case, a perfect $(N,N/2)$-CZCP reduces to a sequence pair, called \textit{strengthened GCP}, whose equivalent CZCP $(\mathbf{c},\mathbf{d})$ (see P1)
is given in (\ref{defi_StreGCP}).
\begin{figure*}
\begin{equation}\label{defi_StreGCP}
\left [
\begin{matrix}
\mathbf{c}\\ \hdashline
\mathbf{d}
\end{matrix}
\right ]=\left [
\begin{array}{ccccc:cccc}
c_0 & c_1 & c_2 &\cdots & c_{N/2-1} & c_{N/2} & c_{N/2+1} & \cdots & c_{N-1}\\ \hdashline
c_0 & c_1 & c_2 &\cdots & c_{N/2-1} & -c_{N/2} & -c_{N/2+1} & \cdots & -c_{N-1}
\end{array}
\right ].
\end{equation}
\end{figure*}
\end{definition}

\vspace{0.1in}
We illustrate P2 by the following example.
\begin{example}
 In the context of the quaternary $(9,3)$-CZCP $(\mathbf{a},\mathbf{b})$ in \textit{Example \ref{exmp_CZCP_len9}}, we have
 {
 \begin{equation}
\underline{\mathbf{b}}^*=\omega_4^{[3,1,1,0,3,0,3,3,0]},~-\underline{\mathbf{a}}^*=\omega_4^{[3,1,1,0,2,0,1,1,2]}.
\end{equation}
}
In addition to $\rho\left(\mathbf{a}\right)(\tau)+\rho\left(\mathbf{b}\right)(\tau)$ and $\rho\left(\mathbf{a},\mathbf{b}\right)(\tau)+\rho\left(\mathbf{b},\mathbf{a}\right)(\tau)$ shown in \textit{Example \ref{exmp_CZCP_len9}}, we have
\begin{equation}
\begin{split}
 & ~\left (\Bigl |\rho\left(\underline{\mathbf{b}}^*\right)(\tau)+\rho\left(-\underline{\mathbf{a}}^*\right)(\tau) \Bigl | \right )_{\tau=0}^{8}\\
 = & ~(18,\mathbf{0}_{1\times3},2\sqrt{2},2,\mathbf{0}_{1\times 3}),\\
 & ~\left (\Bigl |\rho(\mathbf{a},\underline{\mathbf{b}}^*)(\tau)+\rho(\mathbf{b},-\underline{\mathbf{a}}^*)(\tau) \Bigl | \right )_{\tau=0}^{8}\\
 = & ~\left (\mathbf{0}_{1\times 9}\right ),\\
 & ~\left (\Bigl |\rho(\mathbf{b},\underline{\mathbf{b}}^*)(\tau)+\rho(\mathbf{a},-\underline{\mathbf{a}}^*)(\tau) \Bigl | \right )_{\tau=0}^{8}\\
 = & ~(2\sqrt{13},2\sqrt{26},2\sqrt{2},2\sqrt{2},2\sqrt{10},4,\mathbf{0}_{1\times 3}).
\end{split}
\end{equation}
 \end{example}

\vspace{0.1in}
Note that the \textit{even} value constraint on sequence length $N$ in P3 may not be necessary for $q$-ary CZCPs with $q>2$. In other words,
there exist $q$-ary $(N,Z)$-CZCPs with odd $N$ provided that $Z\leq N/2$ (see \textit{Example \ref{exmp_CZCP_len9}}). One can verify (\ref{equ_P1}) in P1 by noting that the first three elements of both $\mathbf{a}$ and $\mathbf{b}$ in \textit{Example \ref{exmp_CZCP_len9}} are pairwisely identical, whereas the last three elements have pairwisely opposite polarities when complex-valued elements are considered.

\vspace{0.1in}
Taking advantage of \textit{Properties 1-3}, we have carried out a computer search for binary (equivalent) $(N,Z)$-CZCPs of lengths up to 26. These search
results are presented in Table I, in which the maximum $Z$ is achieved for every binary CZCP with length $N$. One can see that not all the even $N$ have \textit{perfect} binary CZCPs with $Z=N/2$.

\vspace{0.1in}
\renewcommand{\arraystretch}{0.6}

\begin{table*}[!ht]
\caption{A list of binary $(N,Z)$-CZCPs (with maximum $Z$) of lengths up to 26}
\small
\centering
\tabcolsep=0.01cm
\begin{tabular}{|c||c|c|}
\hline
$(N,Z)$ & $\left (\begin{matrix}
        {\mathbf{a}}\\
        {\mathbf{b}}
        \end{matrix} \right )$  &  $\left ( \begin{matrix}\Bigl |\rho({\mathbf{a}})(\tau)+\rho({\mathbf{b}})(\tau)\Bigl |\\
                                                           \Bigl |\rho({\mathbf{a}},\mathbf{b})(\tau)+\rho({\mathbf{b}},\mathbf{a})(\tau)\Bigl |
                                            \end{matrix}\right )_{\tau=0}^{N-1}$  \\ \hline \hline
$(2,1)$   & $\left( \begin{matrix}
            ++\\
            +-
          \end{matrix} \right) $ & $\left( \begin{matrix}
            4&0\\
            0&0
          \end{matrix} \right) $         \\  \hline
$(4,2)$   & $ \left (\begin{matrix}
        +++-\\
        ++-+
    \end{matrix} \right )$ & $\left( \begin{matrix}
            8&0&\mathbf{0}_2\\
            0&4&\mathbf{0}_2
          \end{matrix} \right) $        \\  \hline
$(6,2)$   & $ \left (\begin{matrix}
        ++++-+\\
        ++-++-
    \end{matrix} \right )$ & $\left( \begin{matrix}
            12&0&0&4&\mathbf{0}_2\\
            0&4&4&0&\mathbf{0}_2
          \end{matrix} \right) $          \\  \hline
$(8,4)$   & $ \left ( \begin{matrix}
        +++-++-+\\
        +++---+-
    \end{matrix} \right )$ & $\left( \begin{matrix}
            16&0&0&0&\mathbf{0}_4\\
            0&4&0&4&\mathbf{0}_4
          \end{matrix} \right) $        \\  \hline
$(10,4)$   & $ \left ( \begin{matrix}
           ++-+++++--\\
           ++-+-+--++
    \end{matrix} \right) $ & $\left( \begin{matrix}
            20&0&0&0&0&0&\mathbf{0}_4\\
            0&4&4&0&4&4&\mathbf{0}_4
          \end{matrix} \right) $        \\  \hline
$(12,5)$   & $ \left ( \begin{matrix}
++++-++--+-+\\
++++--+++-+-
    \end{matrix} \right )$ & $\left( \begin{matrix}
            24&0&0&0&0&0&4&\mathbf{0}_5\\
            0&8&0&4&0&4&0&\mathbf{0}_5
          \end{matrix} \right) $         \\  \hline
$(14,6)$   & $ \left ( \begin{matrix}
+++-+-+++++--+\\
+++-+--+---++-
    \end{matrix} \right )$ & $\left( \begin{matrix}
            28&0&0&0&0&0&0&4&\mathbf{0}_6\\
            0&4&4&0&4&0&4&0&\mathbf{0}_6
          \end{matrix} \right) $         \\  \hline
$(16,8)$   & $ \left ( \begin{matrix}
+++-++-++-+++---\\
+++-++-+-+---+++
    \end{matrix} \right )$ & $\left( \begin{matrix}
            32&0&0&0&0&0&0&0&\mathbf{0}_8\\
            0&4&0&12&0&4&0&4&\mathbf{0}_8
          \end{matrix} \right) $         \\  \hline
$(18,7)$   & $ \left ( \begin{matrix}
++-++++-----++-+-+\\
++-+++++--++--+-+-
    \end{matrix} \right )$ &  $\left( \begin{array}{cccccccccccc}
            36&0 &0&0&0&0&0&0&6&0&2&\mathbf{0}_7\\
             0&12&0&0&4&0&4&0&2&4&2&\mathbf{0}_7
          \end{array} \right) $        \\  \hline
$(20,10)$   & $ \left ( \begin{matrix}
+--++++++-+--+-+---+\\
+--++++++--++-+-+++-
    \end{matrix} \right )$ &$\left( \begin{array}{cccccccccccc}
            40&0 &0&0&0&0&0&0&0&0&\mathbf{0}_{10}\\
             0&12&0&4&0&4&8&4&8&4&\mathbf{0}_{10}
          \end{array} \right) $         \\  \hline
$(22,9)$  & $ \left (\begin{matrix}
++++-+-++---++++-++--+\\
++++-+-+++-+----+--++-
    \end{matrix} \right )$ & $\left( \begin{array}{cccccccccccccc}
            44&0&0&0&0&0&0&0&0&0&2&2&2&\mathbf{0}_{9}\\
             4&0&8&4&0&4&8&4&0&4&2&2&2&\mathbf{0}_{9}
          \end{array} \right) $         \\  \hline
$(24,11)$   & $ \left ( \begin{matrix}
++++++---++--+--+--+-+-+\\
++++++---+++--++-++-+-+-
    \end{matrix} \right )$ &  $\left( \begin{array}{cccccccccccccc}
            48&0&0 &0&0&0&0&0&0&0&0&0&2&\mathbf{0}_{11}\\
            0&24&0&12&0&4&0&4&0&4&0&4&0&\mathbf{0}_{11}
          \end{array} \right) $      \\  \hline
$(26,12)$   & $ \left ( \begin{matrix}
           ++++-++--+-+-+--+-+++--+++\\
           ++++-++--+-+++++-+---++---
    \end{matrix} \right )$ &  $\left( \begin{array}{cccccccccccccccc}
            52&0&0&0&0&0&0&0&0&0&0&0&0&0&\mathbf{0}_{12}\\
            0 &4&4&8&4&8&4&8&4&0&4&8&4&4&\mathbf{0}_{12}
          \end{array} \right) $      \\  \hline
\end{tabular}
\end{table*}

\renewcommand{\arraystretch}{1.0}

\vspace{0.1in}
Next, we present a systematic approach to construct  {\textit{perfect}} CZCPs through \textit{strengthened} GCPs (see \textit{Definition \ref{defi_SGCP}}).
\vspace{0.1in}

 {
\begin{construction}\label{Cons_CZCP}
Let $(\mathbf{e},\mathbf{f})$ be $q$-ary GCP of length $N/2$. Then, every sequence pair (arranged in matrix form with two rows) in (\ref{StreGCP2}) is a  {\textit{perfect}} CZCP (i.e., \textit{strengthened GCP}).
\begin{figure*}
\begin{equation}\label{StreGCP2}
\left\{
\left [
\begin{matrix}
\omega_q^{\upsilon_1} \cdot \mathbf{e}, ~\omega_q^{\upsilon_1+\upsilon}\cdot {\mathbf{f}}\\ \hdashline
\omega_q^{\upsilon_2}\cdot \mathbf{e}, -\omega_q^{\upsilon_2+\upsilon}\cdot {\mathbf{f}}
\end{matrix}
\right ],
\left [
\begin{matrix}
\omega_q^{\upsilon_1} \cdot\mathbf{e}, -\omega_q^{\upsilon_1+\upsilon} \cdot{\mathbf{f}}\\ \hdashline
\omega_q^{\upsilon_2}\cdot \mathbf{e}, ~\omega_q^{\upsilon_2+\upsilon} \cdot {\mathbf{f}}
\end{matrix}
\right ],
\left [
\begin{matrix}
\omega_q^{\upsilon_1} \cdot\mathbf{f}, ~\omega_q^{\upsilon_1+\upsilon} \cdot {\mathbf{e}}\\ \hdashline
\omega_q^{\upsilon_2}\cdot \mathbf{f}, -\omega_q^{\upsilon_2+\upsilon} \cdot {\mathbf{e}}
\end{matrix}
\right ],
\left [
\begin{matrix}
\omega_q^{\upsilon_1}\cdot \mathbf{f}, -\omega_q^{\upsilon_1+\upsilon} \cdot {\mathbf{e}}\\ \hdashline
\omega_q^{\upsilon_2}\cdot \mathbf{f}, ~\omega_q^{\upsilon_2+\upsilon} \cdot {\mathbf{e}}
\end{matrix}
\right ]
\right \},
\end{equation}
\end{figure*}
where $\upsilon_1,\upsilon_2,\upsilon\in \mathbb{Z}_q (q~\text{even})$ and $\upsilon_1-\upsilon_2\in \{0,q/2\}(\text{mod}~q)$.
\end{construction}
\begin{proof}
We only prove the sequence pair of $\mathbf{a}=[\omega_q^{\upsilon_1}\cdot \mathbf{e},\omega_q^{\upsilon_1+\upsilon} \cdot\mathbf{f}],\mathbf{b}=[\omega_q^{\upsilon_2} \cdot \mathbf{e},-\omega_q^{\upsilon_2+\upsilon} \cdot {\mathbf{f}}]$ here. The proofs for other cases
are omitted as they can be obtained in a similar way. It is readily to show that $(\mathbf{a},\mathbf{b})$ is a GCP provided that $(\mathbf{e},\mathbf{f})$
is a GCP \cite{Golay61,Fan-book}. For $N/2\leq \tau\leq N-1$, we have
\begin{equation}
\begin{split}
 &\rho\left(\mathbf{a},\mathbf{b}\right)(\tau)+\rho\left(\mathbf{b},\mathbf{a}\right)(\tau)\\
 = & -\rho\left(\omega_q^{\upsilon_1}\cdot\mathbf{e},\omega_q^{\upsilon_2+\upsilon}\cdot\mathbf{f}\right)(\tau)+\rho\left(\omega_q^{\upsilon_2}\cdot\mathbf{e},\omega_q^ {\upsilon_1+\upsilon}\cdot\mathbf{f}\right)(\tau)\\
 = & \left [ -\omega_q^{\upsilon_1-\upsilon_2}+\omega_q^{\upsilon_2-\upsilon_1}\right ]\omega_q^{-\upsilon}\cdot \rho\left(\mathbf{e},\mathbf{f}\right)(\tau).
\end{split}
\end{equation}
Hence, we assert that the second condition of (\ref{MSE_min_ACCFcond}) is held if $\upsilon_1-\upsilon_2\equiv q/2 (\text{mod}~q)$ or $\upsilon_1-\upsilon_2\equiv 0 (\text{mod}~q)$ .
\end{proof}
}

\vspace{0.1in}

{
\begin{example}
Consider the length-11 quaternary GCP $(\mathbf{e},\mathbf{f})$ below
\begin{displaymath}
\mathbf{e}=\omega_4^{[0,1,2,0,2,1,3,2,1,1,0]},~\mathbf{f}=\omega_4^{[0,0,3,3,3,0,0,1,2,0,2]}.
\end{displaymath}
Let $d=1$. We obtain a \textit{perfect} CZCP (i.e., strengthened GCP) $(\mathbf{a},\mathbf{b})$ as follows.
\begin{equation}\label{optiCZCP_ex}
\begin{array}{cll}
\mathbf{a}&=[\mathbf{e}, \omega_4^d \cdot \mathbf{f}]&=\omega_4^{[0,1,2,0,2,1,3,2,1,1,0, 1,1,0,0,0,1,1,2,3,1,3]},\\
\mathbf{b}&=[\mathbf{e},-\omega_4^d \cdot \mathbf{f}]&=\omega_4^{[0,1,2,0,2,1,3,2,1,1,0, 3,3,2,2,2,3,3,0,1,3,1]}.
\end{array}
\end{equation}
From (\ref{optiCZCP_ex}), we have
\begin{equation}
\begin{split}
 & ~\left (\Bigl |\rho\left(\mathbf{a}\right)(\tau)+\rho\left(\mathbf{b}\right)(\tau) \Bigl | \right )_{\tau=0}^{21}\\
 = & ~(44,\mathbf{0}_{1\times21}),\\
 & ~\left (\Bigl |\rho\left(\mathbf{b},\mathbf{a}\right)(\tau)+\rho\left(\mathbf{a},\mathbf{b}\right)(\tau) \Bigl | \right )_{\tau=0}^{21}\\
 = & ~\left (0,8\sqrt{2},4,4\sqrt{2},4,0,4,4\sqrt{2},4,0,4,\mathbf{0}_{1\times11} \right ).
\end{split}
\end{equation}
\end{example}
}
\vspace{0.1in}
{
\begin{construction}\label{Cons_CZCP_GBF}
In the context of \textit{Lemma \ref{PSK_GDJ constr_4GCP}}, let $\pi(1)=\mu$ and $w'\in \{0,q/2\}$ (\text{mod}~$q$). Then, $\varphi_q(\mathbf{{g}})$ and $\varphi_q(\mathbf{{g}}+\frac{q}{2}{\mathbf{x}}_{\pi(1)}+w'\cdot \mathbf{1})$ form a perfect $q$-ary CZCP of length $2^\mu$.
\end{construction}
\begin{proof}
Let $\mathbf{g}_1,\mathbf{g}_2$, each having length $2^{\mu-1}$, be the first- and second- halves of sequence $\mathbf{g}$, respectively. Similarly, we define $\mathbf{g}'_1,\mathbf{g}'_2$ for that of sequence $\mathbf{{g}}+\frac{q}{2}{\mathbf{x}}_{\pi(1)}+w'\cdot \mathbf{1}$. Hence,
\begin{equation}
\mathbf{g}=[\mathbf{g}_1,\mathbf{g}_2],~\mathbf{{g}}+\frac{q}{2}{\mathbf{x}}_{\pi(1)}+w'\cdot \mathbf{1}=[\mathbf{g}'_1,\mathbf{g}'_2].
\end{equation}
Denote by $g_1,g_2,g'_1,g'_2$ the GBFs corresponding to $\mathbf{g}_1,\mathbf{g}_2,\mathbf{g}'_1,\mathbf{g}'_2$, respectively.
By setting $x_{\pi(1)}=0$ and $1$ into (\ref{f_4GDJ_GCP}), and noting that $\pi(1)=\mu$, we obtain the GBFs corresponding to the first- and second- halves of $\mathbf{g}$, respectively. Specifically, we have 
\begin{equation}
\begin{split}
g_1 & = \frac{q}{2} \sum \limits_{k=2}^{\mu-1}x_{\pi(k)} x_{\pi(k+1)} + \sum
\limits_{k=1}^{\mu-1} w_k x_k+w,\\
g_2 & = g_1+\frac{q}{2}x_{\pi(2)}+w_\mu.
\end{split}
\end{equation}
In a similar way, we have
\begin{equation}
\begin{split}
g'_1 & = g_1+w',\\
g'_2 & = g_1+\frac{q}{2}x_{\pi(2)}+w_\mu+w'+q/2.
\end{split}
\end{equation}
To show that $\varphi_q(\mathbf{{g}})$ and $\varphi_q(\mathbf{{g}}+\frac{q}{2}{\mathbf{x}}_{\pi(1)}+w'\cdot \mathbf{1})$ form a perfect CZCP, we only need to check if the second condition of (\ref{MSE_min_ACCFcond}) can be satisfied. Let us recall \textit{Definition \ref{defi_correGBFs}} for aperiodic correlation of complex-valued sequences associated to GBFs. Consider $2^{\mu-1}\leq \tau<2^\mu$ and then carry out the calculation below.
 \begin{equation}\label{Const2_equ}
 \begin{split}
  & \rho\left(\varphi_q(\mathbf{{g}}),\varphi_q(\mathbf{{g}}+\frac{q}{2}{\mathbf{x}}_{\pi(1)}+w'\cdot \mathbf{1})\right)(\tau)\\
   & ~~~~+\rho\left(\varphi_q(\mathbf{{g}}+\frac{q}{2}{\mathbf{x}}_{\pi(1)}+w'\cdot \mathbf{1}),\varphi_q(\mathbf{{g}})\right)(\tau)\\
 = & \rho_q\left(g,g+\frac{q}{2}x_{\pi(1)}+w'\right)(\tau)+\rho_q\left(g+\frac{q}{2}x_{\pi(1)}+w',g\right)(\tau)\\
 = & \rho_q\left (g_1,\underbrace{g_1+\frac{q}{2}x_{\pi(2)}+w_\mu+w'+q/2}_{=g'_2}\right )(\tau-2^{\mu-1})\\
    & ~~~~~~~~+\rho_q\left (\underbrace{g_1+w'}_{=g'_1},\underbrace{g_1+\frac{q}{2}x_{\pi(2)}+w_\mu}_{=g_2}\right )(\tau-2^{\mu-1})\\
 = & -\omega^{-w'}_q \cdot \rho_q\left (g_1,g_1+\frac{q}{2}x_{\pi(2)}+w_\mu\right )(\tau-2^{\mu-1})\\
    & ~~~+\omega^{w'}_q \cdot \rho_q\left (g_1,g_1+\frac{q}{2}x_{\pi(2)}+w_\mu\right )(\tau-2^{\mu-1})\\
 = & \left [-\omega^{-w'}_q+\omega^{w'}_q\right ]\rho_q\left (g_1,g_1+\frac{q}{2}x_{\pi(2)}+w_\mu\right )(\tau-2^{\mu-1}).
 \end{split}
 \end{equation}
 Clearly, the above equation reduces to zero as $\left [-\omega^{-w'}_q+\omega^{w'}_q\right ]=0$ holds for $w'\in \{0,q/2\}$.
\end{proof}
\vspace{0.1in}

We illustrate \textit{Construction \ref{Cons_CZCP_GBF}} by the example below.

\vspace{0.1in}
\begin{example}
Consider an quaternary CZCP of length-16, i.e., $q=4,\mu=4$. Furthermore, let us set $\pi=[4,2,3,1],[w_1,w_2,w_3,w_4]=[3,2,0,1], w=0, w'=2$. By \textit{Lemma \ref{PSK_GDJ constr_4GCP}}, we obtain a GCP below.
\begin{equation}\label{optiCZCP_ex2}
\begin{array}{cl}
\mathbf{a}&=\omega_4^{[0,     3,     2,     1,     0,     1,     0,     1,     1,     0,     1,     0,     1,     2,     3,     0]},\\
\mathbf{b}&=\omega_4^{[2,     1,     0,     3,     2,     3,     2,     3,     1,     0,     1,     0,     1,     2,     3,     0]}.
\end{array}
\end{equation}
We assert that the above GCP is a perfect CZCP by checking the magnitudes of their AAC sum and ACC sum as follows:
\begin{equation}
\begin{split}
 & ~\left (\Bigl |\rho\left(\mathbf{a}\right)(\tau)+\rho\left(\mathbf{b}\right)(\tau) \Bigl | \right )_{\tau=0}^{15}\\
 = &~(32,\mathbf{0}_{1\times15}),\\
 & ~\left (\Bigl |\rho\left(\mathbf{b},\mathbf{a}\right)(\tau)+\rho\left(\mathbf{a},\mathbf{b}\right)(\tau) \Bigl | \right )_{\tau=0}^{15}\\
 = & ~\left (0,12,0,4,0,4,0,4,\mathbf{0}_{1\times 8} \right ).
\end{split}
\end{equation}
\end{example}
}

\vspace{0.1in}

\begin{remark}
Note that binary GCPs are only known to exist for lengths $2^{\alpha_1} 10^{\alpha_2} 26^{\alpha_3}$ \cite{Fan-book,Parker02}, where $\alpha_1,\alpha_2,\alpha_3$ are non-negative integers
satisfying $\alpha_1+\alpha_2+\alpha_3\geq 1$. By \textit{Construction \ref{Cons_CZCP_GBF}}, one can see that  {\textit{perfect}} CZCPs (i.e., strengthened GCPs) exist for lengths of $2^{(\alpha_1+1)} 10^{\alpha_2} 26^{\alpha_3}$ where $\alpha_1+\alpha_2+\alpha_3\geq 0$.
\end{remark}

\vspace{0.1in}
{\begin{remark}
\textit{Construction \ref{Cons_CZCP_GBF}} may be viewed as a GBF interpretation of \textit{Construction \ref{Cons_CZCP}} for perfect CZCPs having lengths of power of two. By counting all the possible permutations and linear coefficients of the GBF in (\ref{f_4GDJ_GCP}), one can readily show that \textit{Construction \ref{Cons_CZCP_GBF}} produces $2\cdot\frac{(\mu-1)!}{2}\cdot q^{\mu+1}=(\mu-1)!\cdot q^{\mu+1}$ perfect CZCPs.
\end{remark}
}
\vspace{0.1in}

\begin{figure}
  \centering
  \captionsetup{justification=centering}
  \includegraphics[width=3.2in]{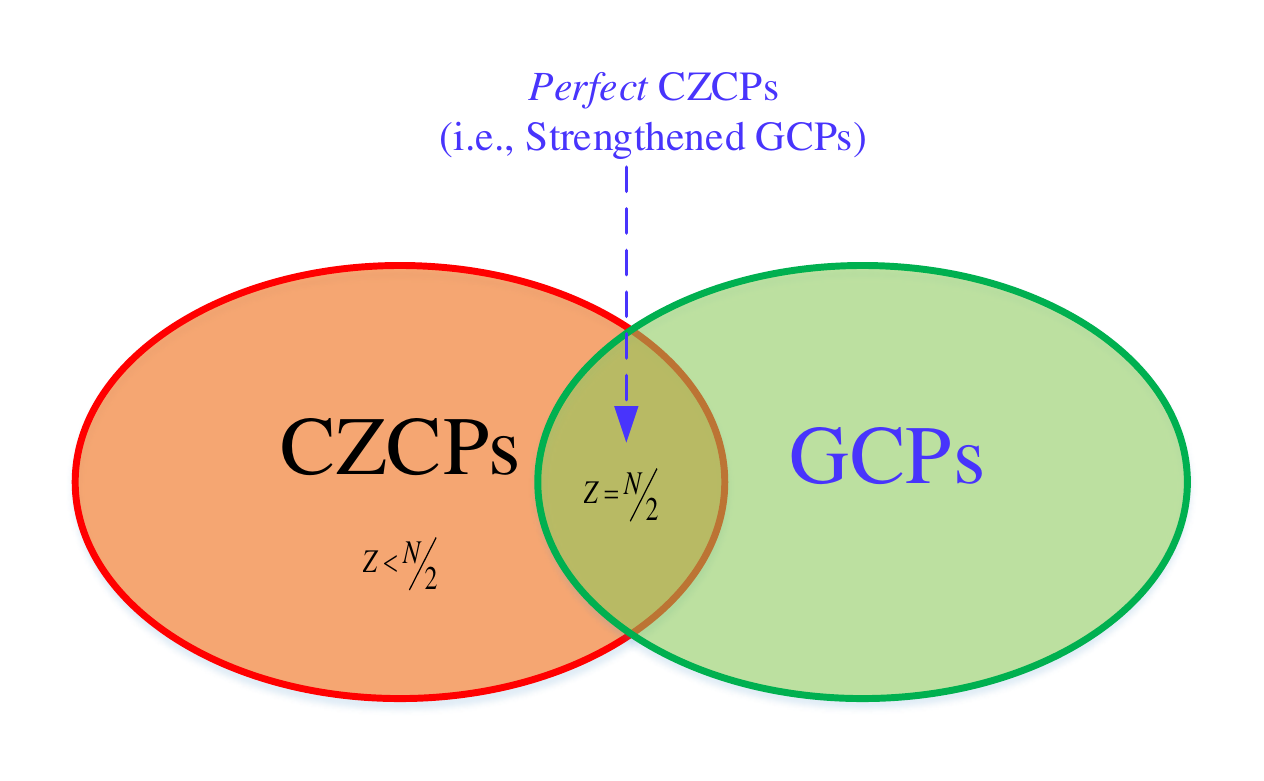}\\
  \caption{Relationship between $(N,Z)$-CZCPs and GCPs.}
  \label{FigCZCPvsGCP}
\end{figure}

We illustrate the relationship between CZCPs and GCPs by Fig. \ref{FigCZCPvsGCP}. It can be seen that CZCPs and GCPs are two different sets of sequence pairs whose intersection is given by  {\textit{perfect}} CZCPs (i.e., strengthened GCPs) with $Z=N/2$. Both CZCPs and GCPs are defined by their aperiodic correlation sums. But CZCPs are different from GCPs in that the former may not necessarily have zero AAC sums for \textit{all} the non-zero time-shifts; instead, they have zero ACC sums for certain time-shifts away from the in-phase position.

{We point it out that the concept of CZCP can be extended to ``cross Z-complementary set (CZCS)" as follows.
\begin{definition}\label{defi_CZCS}
Let $\mathcal{A}=\{\mathbf{a}_1,\mathbf{a}_2,\cdots,\mathbf{a}_M\}$ be a set of sequences of identical length $N$. Recall $\mathcal{T}_1,\mathcal{T}_2$ which are defined in \textit{Definition \ref{CZCP_defi}} based on $N$ and an integer $Z\leq N/2$. $\mathcal{A}$ is called an $(N,Z)$-CZCS if
it possesses \textit{symmetric} zero (out-of-phase) AAC sums for time-shifts over $\mathcal{T}_1\cup \mathcal{T}_2$ and zero ACC sums for time-shifts over $\mathcal{T}_2$. In short, an $(N,Z)$-CZCS needs to satisfy the following two conditions.
\begin{equation}\label{MSE_min_ACCFcond2}
\begin{split}
\text{C1}:&~\sum\limits_{m=1}^{M}\rho\left(\mathbf{a}_m\right)(\tau)  =0,~\text{for all}~|\tau|\in \mathcal{T}_1 \cup \mathcal{T}_2;\\
\text{C2}:&~\sum\limits_{m=1}^{M}\rho\left(\mathbf{a}_m,\mathbf{a}_{\lfloor m+1 \rfloor_{M}}\right)(\tau) =0,~\text{for all}~|\tau| \in \mathcal{T}_2.
\end{split}
\end{equation}
When $M=2$, a CZCS reduces to a CZCP.
\end{definition}
\vspace{0.1in}

 It is noted that an $(N,Z)$-CZCS $\{\mathbf{a}_1,\mathbf{a}_2,\cdots,\mathbf{a}_M\}$ can be readily constructed through an $(N,Z)$-CZCP $(\mathbf{a},\mathbf{b})$. As an example, consider even $M$ and let
 \begin{equation}
 \mathbf{a}_m=
 \begin{cases}
 \mathbf{a}, & m~\text{odd},\\
 \mathbf{b}, & m~\text{even}.
 \end{cases}
 \end{equation}
 It is straightforward to see that the first condition of (\ref{MSE_min_ACCFcond2}) can be met by $\{\mathbf{a}_1,\mathbf{a}_2,\cdots,\mathbf{a}_M\}$. To show the achievability of the second condition of (\ref{MSE_min_ACCFcond2}), let us calculate $\sum_{m=1}^{M}\rho\left(\mathbf{a}_m,\mathbf{a}_{\lfloor m+1 \rfloor_{M}}\right)(\tau)$ as follows:
\begin{equation}
\begin{split}
   & \sum\limits_{m=1}^{M}\rho\left(\mathbf{a}_m,\mathbf{a}_{\lfloor m+1 \rfloor_{M}}\right)(\tau)\\
 = & \sum\limits_{m=1}^{M/2}\rho\left(\mathbf{a}_{2m-1},\mathbf{a}_{2m}\right)(\tau)+\sum\limits_{m=1}^{M/2}\rho\left(\mathbf{a}_{2m},\mathbf{a}_{\lfloor 2m+1 \rfloor_{M}}\right)(\tau)\\
 = & \frac{M}{2}\cdot \left[\rho\left(\mathbf{a},\mathbf{b}\right)(\tau)+\rho\left(\mathbf{b},\mathbf{a}\right)(\tau)\right ]\\
\end{split}
\end{equation}
By recalling the second condition of (\ref{MSE_min_ACCFcond}) associated to $(\mathbf{a},\mathbf{b})$, we complete the proof.}
\vspace{0.1in}

\section{Optimal Training Design Using CZCPs for Broadband SM Systems}
In this section, we study optimal training design for broadband SM systems. Based on a generic training-based SC-MIMO transmission structure, we present an SM training framework using regular sparse matrices and derive the correlation properties of the row sequences of such a sparse matrix. Then, we show that the CZCPs (proposed in Section III) can be utilized as a key component in optimal SM training design.
\subsection{Problem Formulation}
\begin{figure}
  \centering
  \captionsetup{justification=centering}
  \includegraphics[width=3.2in]{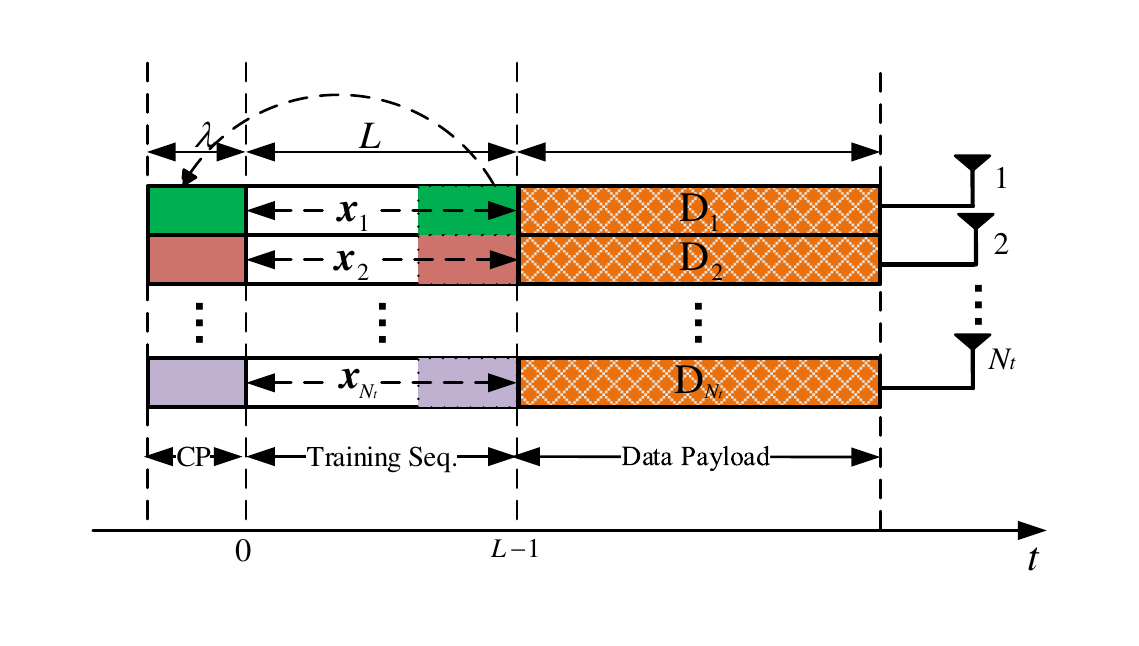}\\
  \caption{Generic training-based SC-MIMO transmission structure with $N_t$ transmit antennas.}
  \label{FigTra4MIMO}
\end{figure}

We consider a generic training-based SC-MIMO transmission structure with $N_t$ TAs as shown in Fig. \ref{FigTra4MIMO}. It subsumes the SC-SM transmission scheme in Fig. \ref{FigSC-SM} as a special case but with emphasis on the training matrix design. We assume a quasi-static frequency-selective channel where the channel impulse response (CIR) from the $n$-th ($1\leq n \leq N_t$) transmit antenna to the receiver is denoted
by $\mathbf{h}_n=[h_{n,0},h_{n,1},\cdots,h_{n,\lambda}]^{\text{T}}$, i.e., a discrete length-$(\lambda+1)$ vector with $h_{n,l}$ ($0\leq l\leq \lambda$) being the
channel coefficient of the $l$-th path. Every block at a TA is divided into two parts: training sequence followed by data payload.
Let $\textit{\textbf{x}}_n=[x_{n,0},x_{n,1},\cdots,x_{n,L-1}]$ be the training sequence transmitted over the $n$-th TA ($1\leq n \leq N_t$). All the training sequences are assumed to have identical energy of $E$, i.e.,
\begin{equation}
\sum\limits_{l=0}^{L-1}|x_{n,l}|^2=E,~~\text{for all}~1\leq n\leq N_t.
\end{equation}
To combat ISI in a frequency-selective channel, a length-$\lambda$ cyclic prefix (CP)\footnote{CP-free MIMO training may be possible with certain sophisticated signal processing algorithm; however, optimal channel estimation performance in frequency-selective channel may not be straightforward due to ISI.} is placed at the front of $\textit{\textbf{x}}_m$ which is comprised of the last $\lambda$ elements of $\textit{\textbf{x}}_n$, i.e., $[x_{n,L-\lambda+2},x_{n,L-\lambda+3},\cdots,x_{n,L-1}]$.

The $k$-th received signal at a RA can be written as
\begin{equation}\label{MIMO_rx_equ}
y_k=\sum\limits_{n=1}^{N_t}\sum\limits_{l=0}^{\lambda}h_{n,l}x_{n,k-l}+w_k,
\end{equation}
where $w_k$ a discrete uncorrelated white complex Gaussian noise sample with zero-mean and variance $\sigma^2_w/2$ per dimension. To proceed, let
\begin{equation}\label{Xm_equ}
\begin{split}
\mathbf{X}_n= &
\left [
\begin{matrix}
x_{n,0} & x_{n,L-1} & \cdots & x_{n,L-\lambda}\\
x_{n,1} & x_{n,0} & \cdots & x_{n,L-\lambda+1}\\
\vdots  & \vdots    & \ddots & \vdots \\
x_{n,L-1} & x_{n,L-2} & \cdots & x_{n,L-\lambda-1}
\end{matrix}
\right ]_{L\times (\lambda+1)},\\
\mathbf{h}= &
\left [
\begin{matrix}
\mathbf{h}_1\\
\mathbf{h}_2\\
\vdots \\
\mathbf{h}_{N_t}
\end{matrix}
\right ]_{N_t (\lambda+1)\times 1},
\end{split}
\end{equation}
and
\begin{equation}
\begin{split}
\mathbf{y} & = [y_0,y_1,\cdots,y_{L-1}]^{\text{T}},\\
\mathbf{X} & = [\mathbf{X}_1,\mathbf{X}_2,\cdots,\mathbf{X}_{N_t}]_{L\times (N_t\lambda+N_t)},\\
\mathbf{w} & = [w_0,w_1,\cdots,w_{L-1}].
\end{split}
\end{equation}
Note that $\mathbf{X}_n$ in (\ref{Xm_equ}) is a Toeplitz matrix in which every column is a cyclically-shifted version of $\textit{\textbf{x}}_n$ and thus has identical energy of $E$. It can be readily shown that (\ref{MIMO_rx_equ}) is equivalent to the following matrix equation.
\begin{equation}
\mathbf{y} = \mathbf{X}\mathbf{h}+\mathbf{w}.
\end{equation}
Applying the least-squares (LS) channel estimator (unbiased), the estimated CIR vector is given by
\begin{equation}
\hat{\mathbf{h}}=\left ( \mathbf{X}^{\text{H}} \mathbf{X}\right )^{-1} \mathbf{X}^{\text{H}} \mathbf{y},
\end{equation}
and the normalized MSE can be derived as
\begin{equation}\label{MSE_equ}
\begin{split}
\text{MSE}&=\frac{1}{(N_t\lambda+N_t)}\text{Tr}\left ( \mathbb{E} \left \{ \Bigl({\mathbf{h}}-\hat{\mathbf{h}}\Bigl)\Bigl({\mathbf{h}}-\hat{\mathbf{h}}\Bigl)^{\text{H}} \right \} \right ) \\
          & =\frac{\sigma^2_w}{(N_t\lambda+N_t)}\text{Tr}\left (  \Bigl(\mathbf{X}^{\text{H}}\mathbf{X}\Bigl)^{-1}  \right ) .
\end{split}
\end{equation}
(\ref{MSE_equ}) indicates that the minimum MSE is achieved if and only if $\mathbf{X}^{\text{H}}\mathbf{X}$ is a diagonal matrix whose elements on the diagonal are identical\footnote{The proof of this statement can be found in the Appendix of \cite{Yang02}.}. Note that
\begin{equation}
\begin{split}
 & \mathbf{X}^{\text{H}}\mathbf{X}\\
 = &
\left [\begin{matrix}
\mathbf{X}_1^{\text{H}}\mathbf{X}_1 & \mathbf{X}_1^{\text{H}}\mathbf{X}_2 & \cdots & \mathbf{X}^{\text{H}}_1\mathbf{X}_{N_t}\\
\mathbf{X}^{\text{H}}_2\mathbf{X}_1 & \mathbf{X}_2^{\text{H}}\mathbf{X}_2 & \cdots & \mathbf{X}^{\text{H}}_2\mathbf{X}_{N_t}\\
\vdots & \vdots & \ddots & \vdots\\
\mathbf{X}^{\text{H}}_{N_t}\mathbf{X}_1 & \mathbf{X}_{N_t}^{\text{H}}\mathbf{X}_2 & \cdots & \mathbf{X}^{\text{H}}_{N_t}\mathbf{X}_{N_t}\\
\end{matrix}
\right ]_{(N_t\lambda+N_t)\times (N_t\lambda+N_t)}.
\end{split}
\end{equation}
Thus, the minimum MSE is achieved if and only if
\begin{equation}\label{MSE_min_cond}
\mathbf{X}^{\text{H}}_i\mathbf{X}_j=
\begin{cases}
E\mathbf{I}_{\lambda+1},~&\text{if}~i=j,\\
\mathbf{0}_{(\lambda+1)\times (\lambda+1)},~&\text{if}~i\neq j,
\end{cases}
\end{equation}
with which we have
\begin{equation}\label{MSE_equ2}
\text{minimum MSE}=\frac{\sigma^2_w}{E}.
\end{equation}
Relating (\ref{MSE_min_cond}) to the PCC between $\textit{\textbf{x}}_i$ and $\textit{\textbf{x}}_j$, we assert that the minimum MSE is achieved if and only if
\begin{equation}\label{MSE_min_cond2}
\phi(\textit{\textbf{x}}_i,\textit{\textbf{x}}_j)(\tau)=
\begin{cases}
E,~~&\text{if}~i=j,\tau=0,\\
0,~~&\text{if}~i\neq j,0\leq \tau \leq \lambda,\\
&~~~~~~~\text{or}~i=j,1\leq \tau\leq \lambda.
\end{cases}
\end{equation}

\vspace{0.1in}
\begin{remark}
From here onwards, (\ref{MSE_min_cond2}) is referred to as the \textit{optimal} condition of SM training sequences under LS channel estimator.
\end{remark}
\vspace{0.1in}

\subsection{Proposed Training Framework For Broadband SM Systems}
To uncover the optimal training sequence criteria for broadband SM systems, we define the following training matrix {\text{\mbox{\boldmath{$\Omega$}}}}.
\begin{equation}\label{TraMatrix_equ}
{\text{\mbox{\boldmath{$\Omega$}}}}=\left [
\begin{matrix}
\textit{\textbf{x}}_1\\
\textit{\textbf{x}}_2\\
\vdots \\
\textit{\textbf{x}}_{N_t}
\end{matrix}
\right ]=
\left [
\begin{matrix}
x_{1,0} & x_{1,1} & \cdots & x_{1,L-1}\\
x_{2,0} & x_{2,1} & \cdots & x_{2,L-1}\\
\vdots  & \vdots    & \ddots & \vdots \\
x_{N_t,0} & x_{N_t,1} & \cdots & x_{N_t,L-1}
\end{matrix}
\right ]_{N_t\times L}.
\end{equation}

\vspace{0.1in}
\begin{remark}
In SM system, only one TA is activated over every time-slot, hence {\text{\mbox{\boldmath{$\Omega$}}}} should be a sparse matrix in which each column has one non-zero entry only (see \textit{Example 1}). This implies that each training sequence $\textit{\textbf{x}}_n$ ($1\leq n \leq N_t$) has $Q\equiv L/N_t$ non-zero entries and $(N_t-1)Q$ zeros.
In this paper, we are concerned with non-zero entries having identical magnitude of 1 (i.e., unimodular) and therefore, each training sequence has energy of $E=Q$.
\end{remark}
\vspace{0.1in}
Suppose that {\text{\mbox{\boldmath{$\Omega$}}}} has the following \textit{sparse} structure.
\begin{equation}
{\text{\mbox{\boldmath{$\Omega$}}}}=\left [
\begin{matrix}
T^0 \left (\mathbf{a}_1\mathbf{0}_{1\times (N_t-1)Q} \right )\\
T^Q \left(\mathbf{a}_2\mathbf{0}_{1\times (N_t-1)Q}  \right )\\
\vdots \\
T^{(N_t-1)Q} \left (\mathbf{a}_{N_t}\mathbf{0}_{1\times(N_t-1)Q} \right)\\
\end{matrix}
\right ]_{N_t\times L},
\end{equation}
where $\mathbf{a}_n=[a_{n,0},a_{n,1},\cdots,a_{n,Q}]$ ($1\leq n\leq N_t$) are row vectors having identical length of $Q$ and energy of $E$. Noted that {\text{\mbox{\boldmath{$\Omega$}}}} is obtained by vertical concatenation of $N_t$ sparse sequences, each of which is a shifted version of a non-zero sequence in $\{\mathbf{a}_1,\mathbf{a}_2,\cdots,\mathbf{a}_{N_t}\}$ padded with $(N_t-1)Q$ zeros.

An example of training matrix {\text{\mbox{\boldmath{$\Omega$}}}} having $N_t=4$ and $L=4Q$ is shown below.
\begin{equation}\label{Omega_examp}
{\text{\mbox{\boldmath{$\Omega$}}}}=\left [
\begin{matrix}
\mathbf{a}_1 & \mathbf{0} & \mathbf{0} & \mathbf{0}\\
\mathbf{0}  & \mathbf{a}_2 & \mathbf{0} & \mathbf{0}\\
\mathbf{0} & \mathbf{0} & \mathbf{a}_3 & \mathbf{0}\\
\mathbf{0} & \mathbf{0} & \mathbf{0} & \mathbf{a}_4
\end{matrix}
\right ]_{4\times 4Q},
\end{equation}
where $\mathbf{0}$ in (\ref{Omega_examp}) denotes $\mathbf{0}_{1\times Q}$. One can easily show that $\phi(\textit{\textbf{x}}_i,\textit{\textbf{x}}_j)(1)$ is non-zero and therefore,
the condition in (\ref{MSE_min_cond2}) can only be satisfied if $\lambda=0$, i.e., flat-fading channels. This means that the training matrix in (\ref{Omega_examp}) cannot achieve minimum channel estimation MSE [see (\ref{MSE_equ2})] in a frequency-selective channel.

To circumvent the above problem, we propose training matrix {\text{\mbox{\boldmath{$\Omega$}}}} which takes the following structure.
\begin{equation}\label{training_matrix_str}
\begin{split}
{\text{\mbox{\boldmath{$\Omega$}}}}  & =  \Bigl[\Omega_1, \Omega_2,  \cdots , \Omega_J \Bigl],\\
\Omega_j  & = \left [
\begin{matrix}
T^0 \left (\mathbf{a}^j_1\mathbf{0}_{1\times (N_t-1)\theta} \right )\\
T^\theta \left(\mathbf{a}^j_2\mathbf{0}_{1\times (N_t-1)\theta}  \right )\\
\vdots \\
T^{(N_t-1)\theta} \left (\mathbf{a}^j_M\mathbf{0}_{1\times (N_t-1)\theta} \right)\\
\end{matrix}
\right ]_{N_t\times N_t\theta},
\end{split}
\end{equation}
 where $1\leq j\leq J,,J\geq2$, each unimodular row vector $\mathbf{a}^j_n=[a^j_{n,0},a^j_{n,1},\cdots,a^j_{n,\theta-1}]$ ($1\leq j\leq J,1\leq n \leq N_t$) has length of $\theta$. Clearly $N_tJ\theta=L$ and $Q=J\theta=L/N_t$. The rationale is that each TA will send $Q$ non-zero entries over $J$ sub-blocks (where $J\geq2$) so that they can work in a cooperative way to enable the resultant training matrix to meet the optimal condition specified in (\ref{MSE_min_cond2}).

We also define matrix {\text{\mbox{\boldmath{$\Psi$}}}}, which is obtained by removing all the zero vectors and cyclic-shift operators in {\text{\mbox{\boldmath{$\Omega$}}}}, as follows.
\begin{equation}
{\text{\mbox{\boldmath{$\Psi$}}}}  =\left [
\begin{matrix}
\mathbf{a}^1_1 & \mathbf{a}^2_1 & \cdots & \mathbf{a}^J_1\\
\mathbf{a}^1_2 & \mathbf{a}^2_2 & \cdots & \mathbf{a}^J_2\\
\vdots & \vdots & \ddots & \vdots\\
\mathbf{a}^1_{N_t} & \mathbf{a}^2_{N_t} & \cdots & \mathbf{a}^J_{N_t}
\end{matrix}
\right ]_{N_t\times Q}.
\end{equation}
One can see that {\text{\mbox{\boldmath{$\Omega$}}}} can be one-to-one mapped to {\text{\mbox{\boldmath{$\Psi$}}}} and vice versa. As such, {\text{\mbox{\boldmath{$\Psi$}}}} is called the \textit{characteristic matrix} of
 the training matrix {\text{\mbox{\boldmath{$\Omega$}}}}. In the sequel, we sometimes write the training matrix and characteristic matrix as $(N_t,J,\theta)$-{\text{\mbox{\boldmath{$\Omega$}}}}
 and $(N_t,J,\theta)$-{\text{\mbox{\boldmath{$\Psi$}}}}, respectively.

\vspace{0.1in}

\begin{example}
Training matrix $(4,2,\theta)$-{\text{\mbox{\boldmath{$\Omega$}}}} for channel estimation in SM with 4 TAs is shown in Fig. \ref{TraingExample2}. This training matrix and its characteristic matrix can be written as follows.
\begin{equation}
\begin{split}
{\text{\mbox{\boldmath{$\Omega$}}}} & =\left [
\begin{array}{cccc:cccc}
\mathbf{a}^1_1 & \mathbf{0} & \mathbf{0} & \mathbf{0} & \mathbf{a}^2_1 & \mathbf{0} & \mathbf{0} & \mathbf{0} \\
\mathbf{0}  & \mathbf{a}^1_2 & \mathbf{0} & \mathbf{0} & \mathbf{0}  & \mathbf{a}^2_2 & \mathbf{0} & \mathbf{0}\\
\mathbf{0} & \mathbf{0} & \mathbf{a}^1_3 & \mathbf{0} & \mathbf{0} & \mathbf{0} & \mathbf{a}^2_3 & \mathbf{0}\\
\mathbf{0} & \mathbf{0} & \mathbf{0} & \mathbf{a}^1_4 & \mathbf{0} & \mathbf{0} & \mathbf{0} & \mathbf{a}^2_4
\end{array}
\right ]_{4\times 8\theta},\\
{\text{\mbox{\boldmath{$\Psi$}}}}  & =\left [
\begin{matrix}
\mathbf{a}^1_1 & \mathbf{a}^2_1 \\
\mathbf{a}^1_2 & \mathbf{a}^2_2 \\
\mathbf{a}^1_3 & \mathbf{a}^2_3 \\
\mathbf{a}^1_4 & \mathbf{a}^2_4
\end{matrix}
\right ]_{4\times 2\theta},
\end{split}
\end{equation}
where $\mathbf{0}$ in {\text{\mbox{\boldmath{$\Omega$}}}} stands for $\mathbf{0}_{1\times \theta}$.
\begin{figure}
  \centering
  \captionsetup{justification=centering}
  \includegraphics[width=3.2in]{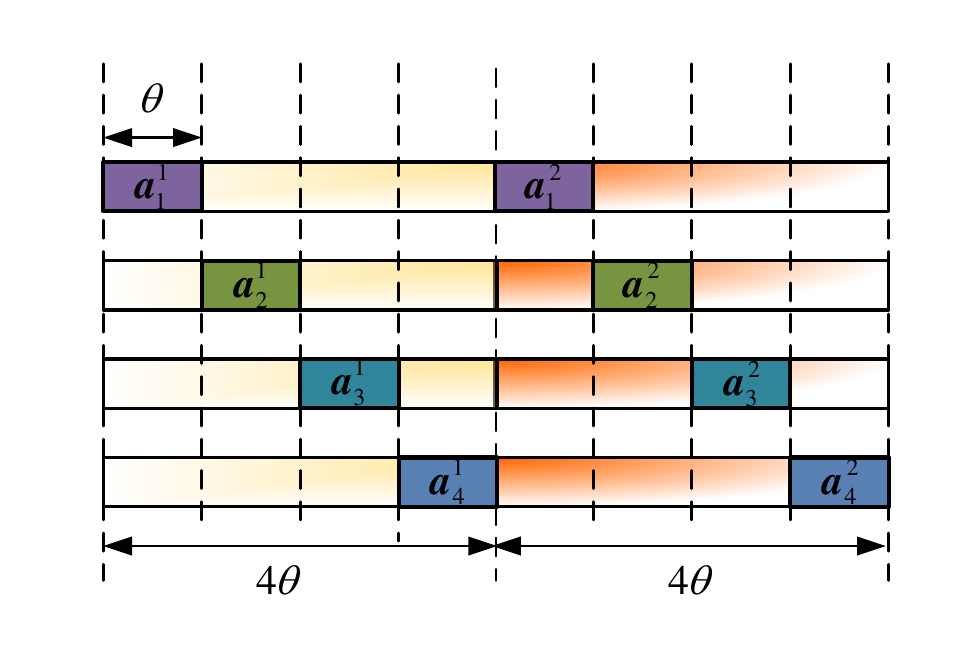}\\
  \caption{Training matrix {\text{\mbox{\boldmath{$\Omega$}}}} with $N_t=4,J=2,L=2N_t\theta$.}
  \label{TraingExample2}
\end{figure}
Note that in frequency-selective channels, the non-zero component sequences in row 1 and row 2 tend to interfere due to multi-path propagation, likewise for row 2 and row 3, row 3 and row 4, row 4 and row 1, so on and so forth. We will show later that such interference can be mitigated by the ``tail-end ZACZ" and ``tail-end ZCCZ" properties of a CZCP (see Fig. \ref{CZCPex}). Also, the ISI of each row can be mitigated by the ``front-end ZACZ" property of CZCP. This makes CZCP a key component in the SM training design.
\end{example}

\vspace{0.1in}

\noindent \textit{Optimal SM Training Design Criteria}: Assume that $\theta\geq \lambda$. Recalling (\ref{MSE_min_cond2}), we assert that the optimal SM training sequences need to satisfy the following three equations\footnote{Note that $\phi(\textit{\textbf{x}}_{i+1},\textit{\textbf{x}}_i)(0)=0,~\forall~1\leq i<N_t$ and $\phi(\textit{\textbf{x}}_{1},\textit{\textbf{x}}_{N_t})(0)=0$ can always be ensured by the proposed training matrix structure in (\ref{training_matrix_str}).}.
\begin{equation}\label{MSE_min_CCFcond1}
\begin{split}
\phi(\textit{\textbf{x}}_{i+1},\textit{\textbf{x}}_i)(\tau) & =\sum\limits_{j=1}^J\rho\left({\mathbf{a}}^j_{i+1},{\mathbf{a}}^j_{i}\right)(\theta-\tau)=0,\\
                                                            & ~1\leq \tau\leq \lambda,1\leq i <N_t,
\end{split}
\end{equation}
\begin{equation}\label{MSE_min_CCFcond2}
\phi(\textit{\textbf{x}}_{1},\textit{\textbf{x}}_{N_t})(\tau)=\sum\limits_{j=1}^J\rho\left({\mathbf{a}}^{\lfloor j \rfloor_J+1}_{1},{\mathbf{a}}^j_{N_t}\right)(\theta-\tau)=0,~1\leq \tau\leq \lambda,
\end{equation}
and
\begin{equation}\label{MSE_min_ACFcond}
\phi(\textit{\textbf{x}}_i)(\tau)= \sum\limits_{j=1}^J\rho\left({\mathbf{a}}^j_{i}\right)(\tau)=0,~1\leq \tau\leq \lambda,1\leq i\leq N_t.
\end{equation}
Note that (\ref{MSE_min_CCFcond1})-(\ref{MSE_min_CCFcond2}) and (\ref{MSE_min_ACFcond}) specify the PCC conditions and PAC condition for the rows of {\text{\mbox{\boldmath{$\Omega$}}}},
respectively. It is assumed that all the non-zero entries of characteristic matrix {\text{\mbox{\boldmath{$\Psi$}}}} are drawn from alphabet set
$\mathcal{A}_q$ (see Subsection II-A for its definition). In particular, we are interested in binary ($q=2$) and
quaternary ($q=4$) characteristic matrices owing to their low implementation
complexity in practice. We have the following remark.

\vspace{0.1in}
\begin{remark}
For any $q$-ary characteristic matrix with even $q$, $J$ should be even. This implies that any binary ($q=2$) or
quaternary ($q=4$) characteristic matrix should have even $J$. 
\end{remark}
\begin{proof}
Setting $\tau=1$ in (\ref{MSE_min_CCFcond1}), we obtain
\begin{equation}\label{MSE_min_CCFcond1_rmk}
\phi(\textit{\textbf{x}}_{i+1},\textit{\textbf{x}}_i)(1)=\sum\limits_{j=1}^J a^j_{i+1,0}\left (a^j_{i,\theta-1}\right )^*.
\end{equation}
It is noted that the right-hand-side of (\ref{MSE_min_CCFcond1_rmk}) is a summation of $J$ entries which are drawn from $\mathcal{A}_q$.
When $q$ is even, every entry $a^j_{i+1,0}\left (a^j_{i,\theta-1}\right )^*$ can only be cancelled by its negation (one of the $J$ entries in $\mathcal{A}_q$) to
ensure $\phi(\textit{\textbf{x}}_{i+1},\textit{\textbf{x}}_i)(1)=0$. Therefore, $J$ must be even.
\end{proof}

\vspace{0.1in}

\noindent \textit{Proposed Construction}: Consider a $(2,2,\theta)$-{\text{\mbox{\boldmath{$\Psi$}}}} ``seed" characteristic matrix as follows.
\begin{equation}\label{seedChaMatrix}
{\text{\mbox{\boldmath{$\Psi$}}}}=\left [
\begin{matrix}
\mathbf{a}^1_1 & \mathbf{a}^2_1 \\
\mathbf{a}^1_2 & \mathbf{a}^2_2
\end{matrix}
\right ]_{2\times 2\theta}.
\end{equation}
Suppose the \textit{optimal} conditions specified in (\ref{MSE_min_CCFcond1})-(\ref{MSE_min_ACFcond}) are satisfied, i.e.,
\begin{equation}\label{MSE_min_CCFcond1_ex3}
\rho\left({\mathbf{a}}^1_{2},{\mathbf{a}}^1_{1}\right)(\theta-\tau)+\rho\left({\mathbf{a}}^2_{2},{\mathbf{a}}^2_{1}\right)(\theta-\tau)=0,~1\leq \tau\leq \lambda,
\end{equation}
\begin{equation}\label{MSE_min_CCFcond2_ex3}
\rho\left({\mathbf{a}}^{2}_{1},{\mathbf{a}}^1_{2}\right)(\theta-\tau)+\rho\left({\mathbf{a}}^{1}_{1},{\mathbf{a}}^2_{2}\right)(\theta-\tau)=0,~1\leq \tau\leq \lambda,
\end{equation}
and
\begin{equation}\label{MSE_min_ACFcond_ex3}
\rho\left({\mathbf{a}}^1_{i}\right)(\tau)+\rho\left({\mathbf{a}}^2_{i}\right)(\tau)=0,~1\leq \tau\leq \lambda,i\in\{1,2\}.
\end{equation}
Based on the $(2,2,\theta)$-{\text{\mbox{\boldmath{$\Psi$}}}} given in (\ref{seedChaMatrix}), we obtain an enlarged $(N_t,2,\theta)$-characteristic matrix ($N_t\geq2$ even) as follows.
\begin{equation}\label{PropChaMatrix}
\begin{split}
\widetilde{{\text{\mbox{\boldmath{$\Psi$}}}}} & =\mathbf{1}_{N_t/2\times 1}\otimes \Psi\\
& =\left [
\begin{array}{c:c}
\mathbf{1}_{N_t/2\times 1}\otimes \mathbf{a}^1_1 & \mathbf{1}_{N_t/2\times 1}\otimes \mathbf{a}^2_1 \\\hdashline
\mathbf{1}_{N_t/2\times 1}\otimes \mathbf{a}^1_2 & \mathbf{1}_{N_t/2\times 1}\otimes \mathbf{a}^2_2
\end{array}
\right ]_{N_t \times 2\theta}.
\end{split}
\end{equation}
The corresponding training matrix of $\widetilde{{\text{\mbox{\boldmath{$\Psi$}}}}}$ can be expressed as
\begin{equation}\label{training_matrix_Omega}
\begin{split}
  \widetilde{{\text{\mbox{\boldmath{$\Omega$}}}}}  = & \left [
\begin{matrix}
\textit{\textbf{x}}_1\\
\textit{\textbf{x}}_2\\
\vdots \\
\textit{\textbf{x}}_{N_t/2}\\ \hdashline
\textit{\textbf{x}}_{N_t/2+1}\\
\vdots \\
\textit{\textbf{x}}_{N_t}
\end{matrix}
\right ]_{N_t\times 2N_t\theta}\\
= & \left [
\begin{array}{c:c}
T^0 \left (\mathbf{a}^1_1\mathbf{0} \right ) & T^0 \left (\mathbf{a}^2_1\mathbf{0} \right )\\
T^\theta \left (\mathbf{a}^1_1\mathbf{0} \right ) & T^\theta \left (\mathbf{a}^2_1\mathbf{0} \right )\\
\vdots & \vdots\\
T^{(N_t/2-1)\theta} \left (\mathbf{a}^1_1\mathbf{0} \right ) & T^{(N_t/2-1)\theta} \left (\mathbf{a}^2_1\mathbf{0} \right )\\ \hdashline
T^{(N_t/2)\theta} \left (\mathbf{a}^1_2\mathbf{0} \right ) & T^{(N_t/2)\theta} \left (\mathbf{a}^2_2\mathbf{0} \right )\\
T^{(N_t/2+1)\theta} \left (\mathbf{a}^1_2\mathbf{0} \right ) & T^{(N_t/2+1)\theta} \left (\mathbf{a}^2_2\mathbf{0} \right )\\
\vdots & \vdots\\
T^{(N_t-1)\theta} \left (\mathbf{a}^1_2\mathbf{0} \right ) & T^{(N_t-1)\theta} \left (\mathbf{a}^2_2\mathbf{0} \right )\\
\end{array}
\right ]_{N_t\times 2N_t\theta},
\end{split}
\end{equation}
where $\mathbf{0}$ stands for $\mathbf{0}_{1\times(N_t-1)\theta}$.
One can see that the \textit{optimal} conditions in (\ref{MSE_min_CCFcond2}) and (\ref{MSE_min_ACFcond}) can be satisfied by (\ref{MSE_min_CCFcond2_ex3}) and (\ref{MSE_min_ACFcond_ex3}), respectively. Furthermore, by (\ref{MSE_min_CCFcond1_ex3}), we have
\begin{equation}
\begin{split}
 &\phi(\textit{\textbf{x}}_{N_t/2+1},\textit{\textbf{x}}_{N_t/2})(\tau)\\
 =&\rho\left({\mathbf{a}}^1_{2},{\mathbf{a}}^1_{1}\right)(\theta-\tau)+\rho\left({\mathbf{a}}^2_{2},{\mathbf{a}}^2_{1}\right)(\theta-\tau)\\
 =&0,~1\leq \tau \leq  \lambda.
 \end{split}
\end{equation}
To ensure that (\ref{MSE_min_CCFcond1}) is satisfied for all the $i$, we also require
\begin{displaymath}
\phi(\textit{\textbf{x}}_{i+1},\textit{\textbf{x}}_{i})(\tau)=0,~\forall~i\in\{1,2,\cdots,N_t-1\}\setminus \{N_t/2\},~1\leq \tau\leq \lambda.
 \end{displaymath}
This leads to the following \textit{additional} {conditions} which should be met by the $(2,2,\theta)$-{\text{\mbox{\boldmath{$\Psi$}}}} ``seed" characteristic matrix:
\begin{equation}\label{MSE_min_ACFcond_ex3_add1}
\begin{split}
 \phi(\textit{\textbf{x}}_{i+1},\textit{\textbf{x}}_{i})(\tau)=&~\rho\left({\mathbf{a}}^1_{1}\right)(\theta-\tau)+\rho\left({\mathbf{a}}^2_{1}\right)(\theta-\tau)\\
 =&~0,~1\leq i \leq N_t/2-1,1\leq \tau \leq  \lambda.
 \end{split}
\end{equation}
\begin{equation}\label{MSE_min_ACFcond_ex3_add2}
\begin{split}
 \phi(\textit{\textbf{x}}_{i+1},\textit{\textbf{x}}_{i})(\tau)=&~\rho\left({\mathbf{a}}^1_{2}\right)(\theta-\tau)+\rho\left({\mathbf{a}}^2_{2}\right)(\theta-\tau)\\
 =&~0,~N_t/2+1\leq i \leq N_t-1,1\leq \tau \leq  \lambda.
 \end{split}
\end{equation}

\vspace{0.1in}
Let $(\mathbf{a},\mathbf{b})$ be an $(N=\theta,Z=\lambda)$-CZCP which is proposed in Section III. Consider two ``seed" characteristic matrices taking the following structures:
\begin{equation}\label{seedmatrices}
{\text{\mbox{\boldmath{$\Psi$}}}}_1=\left [
\begin{matrix}
\mathbf{a} & \mathbf{b} \\
\mathbf{a} & \mathbf{b}
\end{matrix}
\right ]_{2\times 2\theta},~~{\text{\mbox{\boldmath{$\Psi$}}}}_2=\left [
\begin{matrix}
\mathbf{a} & \mathbf{b} \\
\underline{\mathbf{b}}^* & -\underline{\mathbf{a}}^*
\end{matrix}
\right ]_{2\times 2\theta}.
\end{equation}
One can readily show that (\ref{MSE_min_CCFcond1_ex3})-(\ref{MSE_min_ACFcond_ex3}) and (\ref{MSE_min_ACFcond_ex3_add1})-(\ref{MSE_min_ACFcond_ex3_add2})
can be satisfied by both ${\text{\mbox{\boldmath{$\Psi$}}}}_1$ and ${\text{\mbox{\boldmath{$\Psi$}}}}_2$. The resultant ``seed" characteristic matrices
will allow us to design training matrices $(N_t,2,\theta)$-$\widetilde{{\text{\mbox{\boldmath{$\Omega$}}}}}$ [see (\ref{training_matrix_Omega})] for \textit{optimal} channel
estimation in SM system with $N_t$ TAs.

Based on $(N_t,2,\theta)\text{-}\widetilde{{\text{\mbox{\boldmath{$\Omega$}}}}}$, we can also construct a training matrix $(N_t,J,\theta)$-$\overline{{\text{\mbox{\boldmath{$\Omega$}}}}}$ (with longer row sequences and hence larger value of $E$ and enhanced channel estimation performance) by the following expansion rule.
\begin{equation}\label{SMTrain-equ}
\overline{{\text{\mbox{\boldmath{$\Omega$}}}}}=\mathbf{1}_{1\times J/2}\otimes \widetilde{{\text{\mbox{\boldmath{$\Omega$}}}}},~\text{for}~J>2.
\end{equation}

Finally, based on (\ref{seedChaMatrix}), (\ref{PropChaMatrix}), (\ref{training_matrix_Omega}), (\ref{SMTrain-equ}), we summarize the proposed SM training design as follows.
\begin{equation}
\begin{split}
 & (\theta,\lambda)\text{-CZCP} \mapsto (2,2,\theta)\text{-}{\text{\mbox{\boldmath{$\Psi$}}}}\mapsto (N_t,2,\theta)\text{-}\widetilde{{\text{\mbox{\boldmath{$\Psi$}}}}} \\
 & ~~~\mapsto (N_t,2,\theta)\text{-}\widetilde{{\text{\mbox{\boldmath{$\Omega$}}}}} \mapsto (N_t,J,\theta)\text{-}\overline{{\text{\mbox{\boldmath{$\Omega$}}}}}.
\end{split}
\end{equation}
One may choose $(N_t,2,\theta)\text{-}\widetilde{{\text{\mbox{\boldmath{$\Omega$}}}}}$ or $(N_t,J,\theta)\text{-}\overline{{\text{\mbox{\boldmath{$\Omega$}}}}}$ ($J>2$) as the SM training matrix depends on the channel estimation MSE requirement which is determined by $E=J\theta$ [see (\ref{MSE_equ2})].
 \vspace{0.1in}

\subsection{Numerical Evaluation}
In this subsection, we evaluate the proposed SM training sequences over frequency-selective channels. Throughout this subsection, we assume $N_t=4,N_r=1$. Consider a $(\lambda+1)$-path channel (separated by integer symbol durations) having uniform power delay profile as follows:
\begin{equation}
h[t]=\sum_{n=0}^{\lambda}h_i\delta[t-nT],
\end{equation}
where $h_i$'s are complex-valued Gaussian random variables with zero mean and $\mathbb{E}(|h_i|^2)=1$.

\begin{figure*}
\centerline{
\subfloat[No. of multi-paths = 5, $L=32J,J\in\{2,6,18\},Z=4$.]{\includegraphics[trim=3.25cm 9.6cm 3.35cm 10.025cm, clip=true, width=3.2in]{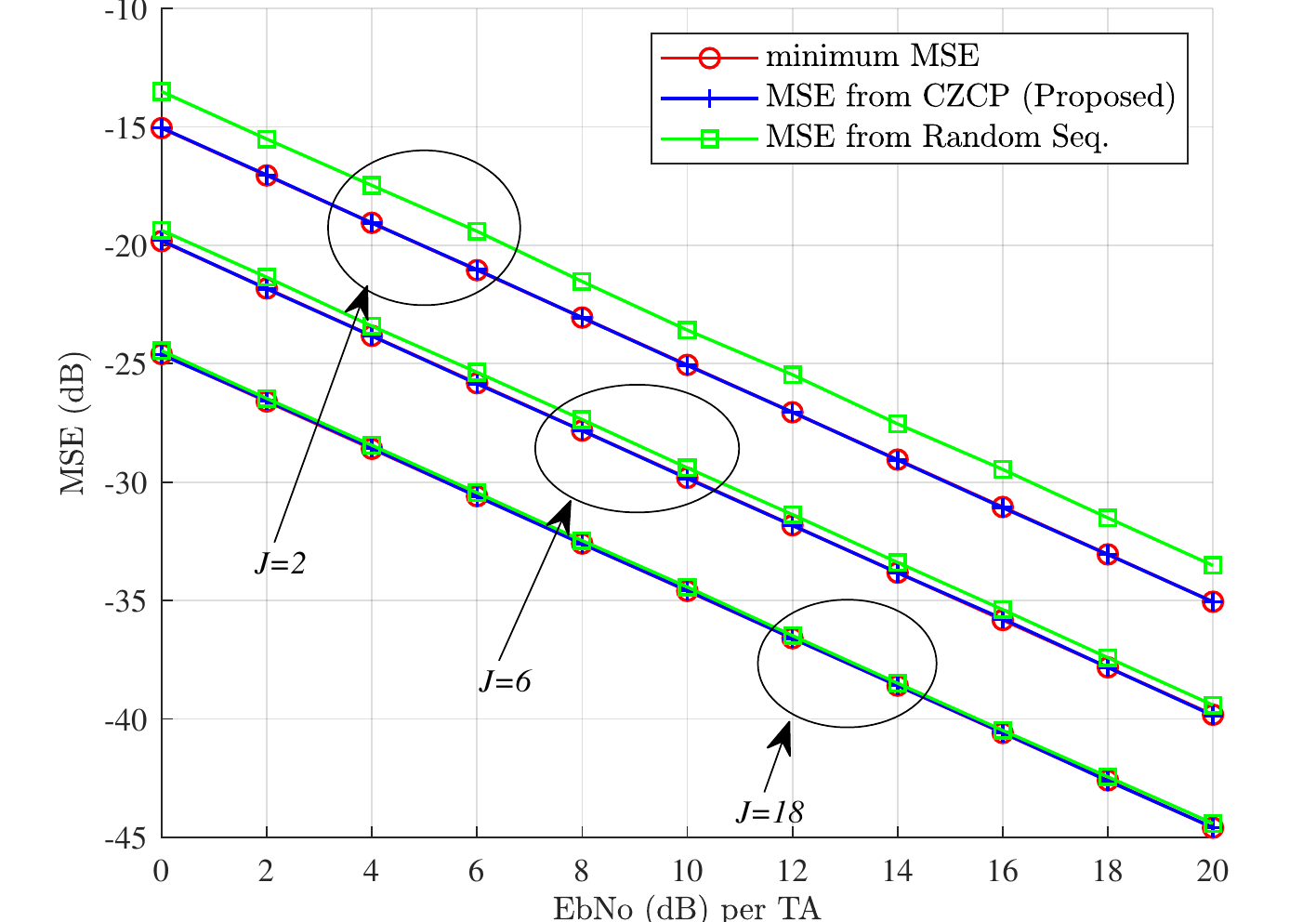}
\label{fig_ex2_a}}
\hspace{-0.1in}
\subfloat[EbNo=16 dB and $E=32$.]{\includegraphics[trim=3.35cm 9.6cm 3.25cm 10.025cm, clip=true, width=3.2in]{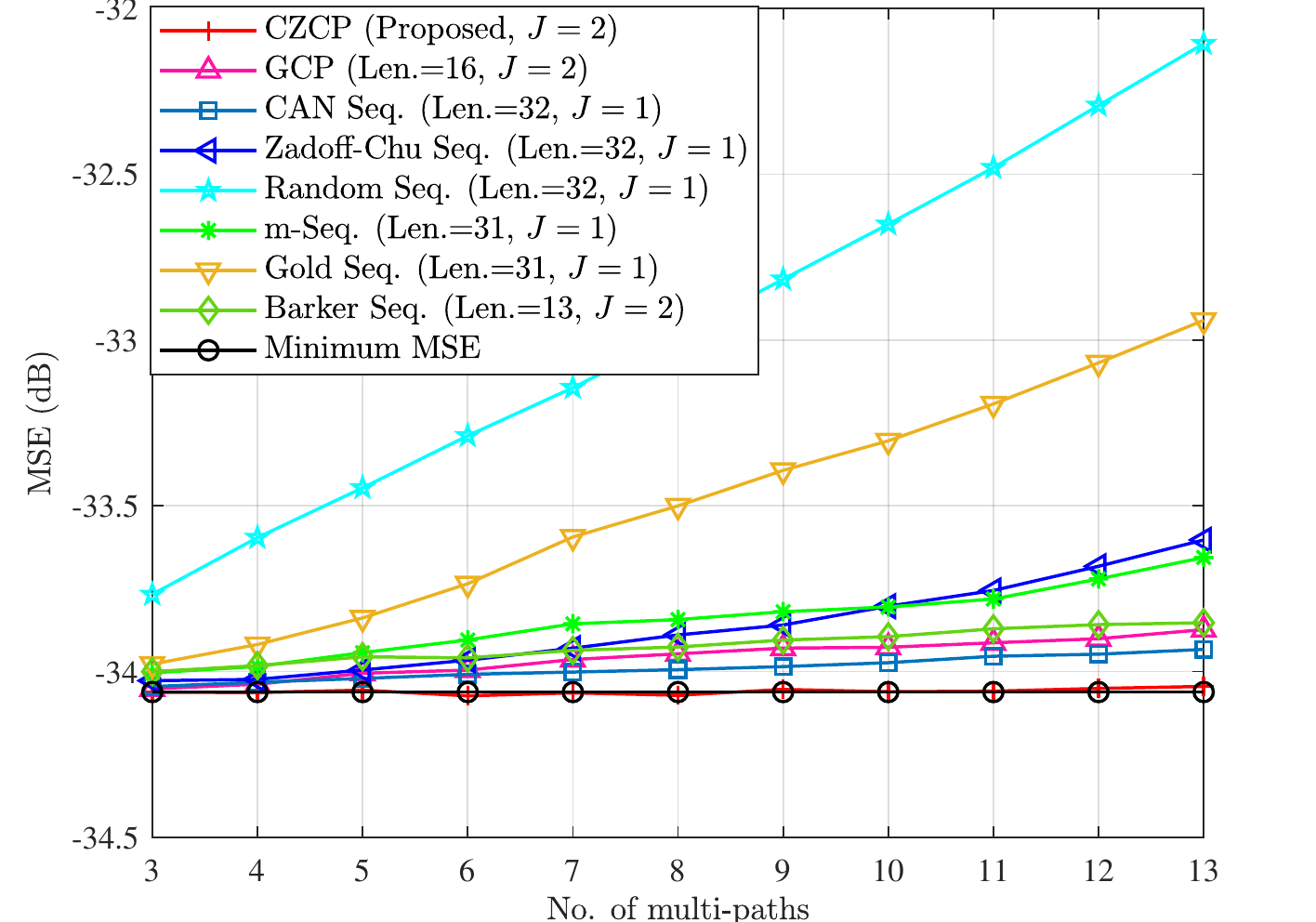}
\label{fig_ex2_b}}
}
\caption{MSE comparison with training matrices from other sequences. }
\label{MSEComp}
\end{figure*}

Let us consider the  {\textit{perfect}} $(N=8,Z=4)$-CZCP given in Table I. As we will see, $(\mathbf{a},\mathbf{b})$ leads to optimal SM training matrix provided that the number of multi-paths is not greater than $Z+1$ (i.e., 5). Based on the training matrix expansion rule in (\ref{training_matrix_Omega}) and applying seed characteristic matrices ${\text{\mbox{\boldmath{$\Psi$}}}}_1$ and ${\text{\mbox{\boldmath{$\Psi$}}}}_2$ given in (\ref{seedmatrices}), respectively, we obtain training matrices $(4,2,8)$-$\widetilde{{\text{\mbox{\boldmath{$\Omega$}}}}}_1$ and $(4,2,8)$-$\widetilde{{\text{\mbox{\boldmath{$\Omega$}}}}}_2$ as follows.
\begin{equation}\label{train_mtx_exam}
\begin{split}
\widetilde{{\text{\mbox{\boldmath{$\Omega$}}}}}_1 & =\left [
\begin{array}{cccc:cccc}
\mathbf{a} & \mathbf{0} & \mathbf{0} & \mathbf{0} & \mathbf{b} & \mathbf{0} & \mathbf{0} & \mathbf{0} \\
\mathbf{0}  & \mathbf{a} & \mathbf{0} & \mathbf{0} & \mathbf{0}  & \mathbf{b} & \mathbf{0} & \mathbf{0}\\
\mathbf{0} & \mathbf{0} & \mathbf{a} & \mathbf{0} & \mathbf{0} & \mathbf{0} & \mathbf{b} & \mathbf{0}\\
\mathbf{0} & \mathbf{0} & \mathbf{0} & \mathbf{a} & \mathbf{0} & \mathbf{0} & \mathbf{0} & \mathbf{b}
\end{array}
\right ],\\
\widetilde{{\text{\mbox{\boldmath{$\Omega$}}}}}_2 & =\left [
\begin{array}{cccc:cccc}
\mathbf{a} & \mathbf{0} & \mathbf{0} & \mathbf{0} & \mathbf{b} & \mathbf{0} & \mathbf{0} & \mathbf{0} \\
\mathbf{0}  & \mathbf{a} & \mathbf{0} & \mathbf{0} & \mathbf{0}  & \mathbf{b} & \mathbf{0} & \mathbf{0}\\
\mathbf{0} & \mathbf{0} & \underline{\mathbf{b}}^* & \mathbf{0} & \mathbf{0} & \mathbf{0} & -\underline{\mathbf{a}}^* & \mathbf{0}\\
\mathbf{0} & \mathbf{0} & \mathbf{0} & \underline{\mathbf{b}}^* & \mathbf{0} & \mathbf{0} & \mathbf{0} & -\underline{\mathbf{a}}^*
\end{array}
\right ],
\end{split}
\end{equation}
where $\mathbf{0}$ in (\ref{train_mtx_exam}) stands for $\mathbf{0}_{1\times 8}$. Taking one SM training matrix in (\ref{train_mtx_exam}) and following the expansion rule in (\ref{SMTrain-equ}), we can obtain $(4,6,8)$-$\overline{{\text{\mbox{\boldmath{$\Omega$}}}}}$ and $(4,18,8)$-$\overline{{\text{\mbox{\boldmath{$\Omega$}}}}}$.
As a comparison, we consider binary \textit{random} (but regular and sparse) training matrices in which each row (of a training matrix) contains $2\lambda J$, where $J\in\{2,6,18\}$, non-zero entries and each column consists of one non-zero entry only. Note that a larger value of $J$ leads to higher value of training sequence energy $E$ and hence better channel estimation performance\footnote{ {In practice, however, a longer training sequence will give rise to a higher training overhead. Hence, selection of the training length (determined by $J$) is a trade-off between channel estimation performance and training overhead.}}. A new random training matrix is generated for every different SM channel estimation (i.e., random on-the-fly). Following Section IV, LS channel estimator is employed. Fig. \ref{MSEComp}-a compares the channel estimation MSEs versus ``EbNo (dB) per TA" for the above two types of training sequences (where $J=2,6,18$) with frequency-selective fading channel consisting of 5 multi-paths. It is seen that the MSE curves obtained from the proposed SM training sequences match with the minimum MSE curves very well for different values of $J$. Compared to the MSE curve obtained from random training sequences, about 1.5 dB gain is achieved for $J=2$. When $J$ increases, the MSE gap between these types of sequences diminishes. This is because the diagonal elements (i.e., taking identical value of $E=8J$) of $\mathbf{X}^{\text{H}}\mathbf{X}$ in (\ref{MSE_equ}) become significantly larger than the off-diagonal ones. Consequently, $\mathbf{X}^{\text{H}}\mathbf{X}$ approaches to an identity matrix as the proposed SM training sequences can achieve.


{Next, we evaluate the channel estimation MSE performances under different values of multi-paths at EbNo of 16 dB. We employ the  {\textit{perfect}} $(N=16,Z=8)$-CZCP given in Table I to generate our proposed SM training matrix with the same structure as $\widetilde{{\text{\mbox{\boldmath{$\Omega$}}}}}_1$ in (\ref{train_mtx_exam}), i.e., $J=2$. We compare its channel estimation performance with SM training matrices from the length-16 GCP  (which is not a CZCP)
\begin{displaymath}
\left [
\begin{matrix}
     1,     1,     1,     1,     1,    -1 ,   -1  ,  1  ,  1  ,  1   ,-1   , -1, 1  , -1  ,  1  , -1\\
     1,    -1,     1,    -1,     1 ,    1 ,   -1 ,  -1  ,   1  ,  -1 ,  -1 ,   1 ,1  ,  1  ,   1 ,    1
\end{matrix}
\right ],
\end{displaymath}
 the length-31 m-sequence\footnote{An m-sequence is a pseudo-random sequence whose periodic autocorrelation sidelobes take identical value of $-1$.}
\begin{displaymath}
\begin{split}
\mathbf{a}_{\text{m}} & =[1  ,  -1 ,   -1 ,   -1 ,   -1 ,    1 ,   -1 ,   -1  ,   1 ,   -1  ,   1 ,    1  ,  -1 ,   -1  ,   1 ,\\
    & ~~~~1  ,   1  ,   1 ,    1 ,   -1 ,   -1,    -1 ,    1 ,    1 ,-1   ,  1   ,  1  ,   1  ,  -1  ,   1 ,   -1],
\end{split}
\end{displaymath}
one length-32 CAN sequence\footnote{CAN sequences are obtained by minimizing a quadratic function associated with the integrated sidelobe level (ISL) of sequences \cite{Stoica2009}. With the aid of fast Fourier transform (FFT), CAN sequences can be generated for lengths up to $10^6$ or even longer. More sequence synthesis algorithms based on numerical optimization can be found in \cite{Stoica2012,Stoica2014,Song2015,Sun2017,Kerahroodi2017,Kerahroodi2019c,Kerahroodi2019j}.} with low aperiodic autocorrelations, the length-13 Barker sequence\footnote{A Barker sequence has aperiodic autocorrelation sidelobes not exceeding 1. It is widely conjectured that no binary Barker sequences exist for lengths larger than 13 \cite{Barker2008} and hence the application of Barker sequences may not be straightforward when a large sequence length is required.}
 \begin{displaymath}
 \mathbf{a}_{\text{B}}=[1, 1, 1, 1, 1, -1, -1, 1, 1, -1, 1, -1, 1],
 \end{displaymath}
four length-31 Gold sequences \cite{Gold1968} with low periodic cross-correlations, four length-32 Zadoff-Chu sequences \cite{Chu1972} with low periodic cross-correlations, and four ``random-on-the-fly" binary sequences of length-32. For the m-sequence, Gold sequences, Zadoff-Chu sequences, CAN sequence and random sequences, we adopt the training matrix structure in (\ref{Omega_examp}). In particular, when a single sequence (i.e., m-sequence or CAN sequence) is used as the ``seed", we set $\mathbf{a}_1=\mathbf{a}_2=\mathbf{a}_3=\mathbf{a}_4$ in (\ref{Omega_examp}).
For the length-13 Barker sequence, we adopt the following training matrix
\begin{displaymath}
\left [
\begin{array}{cccc:cccc}
 \mathbf{a}_{\text{B}} & \mathbf{0} & \mathbf{0} & \mathbf{0} &  \mathbf{a}_{\text{B}} & \mathbf{0} & \mathbf{0} & \mathbf{0} \\
\mathbf{0}  &  \mathbf{a}_{\text{B}} & \mathbf{0} & \mathbf{0} & \mathbf{0}  &  \mathbf{a}_{\text{B}} & \mathbf{0} & \mathbf{0}\\
\mathbf{0} & \mathbf{0} &  \mathbf{a}_{\text{B}} & \mathbf{0} & \mathbf{0} & \mathbf{0} &  \mathbf{a}_{\text{B}} & \mathbf{0}\\
\mathbf{0} & \mathbf{0} & \mathbf{0} &  \mathbf{a}_{\text{B}} & \mathbf{0} & \mathbf{0} & \mathbf{0} &  \mathbf{a}_{\text{B}}
\end{array}
\right ]_{4\times 104}.
\end{displaymath}
For fair comparison, we normalize the energies of all the training sequences to $E=32$. When the number of multi-paths is 9 (or less), $\mathbf{X}^{\text{H}}\mathbf{X}$ corresponding to our proposed SM training matrix is an identity matrix multiplied by the training matrix energy and therefore, our proposed SM training matrix achieves the minimum MSEs, as shown in Fig. \ref{MSEComp}-b. The same may not be achieved by training matrices from other seed sequences. As a result, their channel estimation MSEs display certain distances to the minimum MSE regardless the number of multi-paths. It is interesting to note that the second best channel estimation performance is achieved by the SM training matrix from the length-32 CAN sequence (even better than that from the length-13 Barker sequence) whose aperiodic ISL is minimized. This makes sense because a sequence with lower aperiodic ISL tends to give rise to smaller magnitudes for the off-diagonal elements of $\mathbf{X}^{\text{H}}\mathbf{X}$ and hence better channel estimation MSEs. The worst channel estimation performance is achieved by the SM training matrices from random sequences as it is in general hard to ensure small off-diagonal elements of $\mathbf{X}^{\text{H}}\mathbf{X}$. The second worst channel estimation performance is achieved by the SM training matrices from Gold sequences. Moreover, the channel estimation MSEs corresponding to our proposed SM training matrix start to deviate from, but very close to, the minimum MSE when the number of multi-paths increases to beyond 9. In summary, our proposed SM training matrices give rise to the minimum channel estimation MSEs provided that the number of multi-paths is no greater than $Z+1$; when greater than $Z+1$, our proposed SM training matrices may still give rise to good channel estimation MSEs in the condition that 1) $\mathbf{X}^{\text{H}}\mathbf{X}$ associated to the SM training matrix has full rank and 2) the off-diagonal elements of $\mathbf{X}^{\text{H}}\mathbf{X}$ have small magnitudes. 
}

\section{Conclusions}
In this paper, we have proposed a new class of sequence pairs called cross-complementary pairs (CZCPs) in which every CZCP displays zero (symmetrical) aperiodic auto-correlation (AAC) sums and zero aperiodic cross-correlation (ACC) sums for certain time-shifts. Unlike Golay complementary pairs (GCPs) whose two constituent sequences must be sent over two non-interfering channels (thus only AAC sums matter in GCPs),  {\textit{perfect}} CZCPs should be designed to minimize both ISI (determined by AAC sums) and cross-interference (by ACC sums). By investigating the structural properties of CZCPs, we have found that  {\textit{perfect}} CZCPs are equivalent to a subset of GCPs whose first halves are identical and second halves have opposite polarities (See P2 in Section III). Based on this finding, we have shown by systematic constructions that {\textit{perfect}} binary CZCPs exist for lengths $2^{\alpha+1}10^\beta26^\gamma$, where $\alpha,\beta,\gamma$ are non-negative integers. In \textit{Definition \ref{defi_CZCS}}, we have extended the concept of CZCP to cross Z-complementary set (CZCS) which consists of constituent sequences of two or more.

Secondly, we have shown that CZCPs play an important role in channel training sequence design of broadband spatial modulation (SM) systems. It should be noted that the existing dense training sequences for conventional MIMO systems are unapplicable in SM systems as only one transmit antenna (TA) is activated at each time-slot. By employing CZCP, we have presented a generic framework for the design of \textit{optimal} SM training matrix. We have shown that these training matrices lead to minimum channel estimation mean-squared-error in quasi-static frequency-selective channels.

As a future work, more systematic constructions (e.g., recursive expansion algorithms) for CZCPs/CZCSs (perfect or non-perfect) may be designed. When a {\textit{perfect}} CZCP/CZCS cannot be obtained, it will be interesting to know what are the almost-perfect CZCPs/CZCSs with respect to different sequence lengths and alphabet sizes. Efficient training design for generalized SM \cite{GSM-1,GSM-2,GSM-3} in dispersive channels is also an interesting and challenging issue to be explored. In particular, it is worthy to investigate efficient training sequence design when the generalized SM channels are correlated \cite{Liu2007,Katselis2008,Biguesh2009,Bjornson2010,Shariati2014}.

\section*{Acknowledgment}
The authors are deeply indebted to Prof. Pingzhi Fan at Southwest Jiaotong University for suggesting the terminology of ``CZCP". They also would like to thank anonymous reviewers for their invaluable suggestions which greatly help improve the quality of this work.




\appendices

\section{Proofs of CZCP Properties}

\noindent \textit{Proof of P1}:
\begin{proof}
Since $(\mathbf{c},\mathbf{d})$ is obtained by respectively dividing $\mathbf{a}$ by $a_0$ and $\mathbf{b}$ by
$b_0$, we have $\mathbf{c}=\mathbf{a}/a_0,\mathbf{d}=\mathbf{b}/b_0$. Therefore,
\begin{equation}
\rho(\mathbf{c})(\tau)+\rho(\mathbf{d})(\tau)=\frac{1}{|a_0|^2}\rho(\mathbf{a})(\tau)+\frac{1}{|b_0|^2}\rho(\mathbf{b})(\tau).
\end{equation}
For $q$-ary CZCP $(\mathbf{a},\mathbf{b})$, $a_0,b_0\in\mathcal{A}_q$ and hence $|a_0|^2=|b_0|^2=1$. Thus,
\begin{equation}\label{equ_P1_proof}
\rho(\mathbf{c})(\tau)+\rho(\mathbf{d})(\tau)=\rho(\mathbf{a})(\tau)+\rho(\mathbf{b})(\tau).
\end{equation}
Furthermore, since $\rho(\mathbf{a})(N-1)+\rho(\mathbf{b})(N-1)=\rho(\mathbf{a},\mathbf{b})(N-1)+\rho(\mathbf{a},\mathbf{b})(N-1)=0$, we
have
\begin{equation}
\begin{split}
a_0a^*_{N-1}+b_0b^*_{N-1} & =0,\\
a_0b^*_{N-1}+b_0a^*_{N-1} & =0.
\end{split}
\end{equation}
Multiplying the above two identities by $b^*_0,a^*_0$, respectively, we assert that $a_0b^*_0=a^*_0b_0$ which is a real number. As a result, we have
\begin{equation}\label{equ2_P1_proof}
\begin{split}
\rho(\mathbf{c},\mathbf{d})(\tau)+\rho(\mathbf{d},\mathbf{c})(\tau)& =\frac{1}{b^*_0a_0}\rho(\mathbf{a},\mathbf{b})(\tau)+\frac{1}{b_0a^*_0}\rho(\mathbf{b},\mathbf{a})(\tau)\\
                                                                   & =\frac{1}{b_0a^*_0}\Bigl [\rho(\mathbf{a},\mathbf{b})(\tau)+\rho(\mathbf{b},\mathbf{a})(\tau)\Bigl ].
\end{split}
\end{equation}
Recalling the two conditions specified in (\ref{MSE_min_ACCFcond}), (\ref{equ_P1_proof}) and (\ref{equ2_P1_proof}) indicate that $(\mathbf{c},\mathbf{d})$ is also $q$-ary $(N,\lambda)$-CZCP (with $c_0=d_0=1$).

By $\rho(\mathbf{c})(N-1)+\rho(\mathbf{d})(N-1)=0$, we have
\begin{equation}
c_{N-1}+d_{N-1}=0.
\end{equation}
Based on this (i.e., $c_0=d_0,c_{N-1}=-d_{N-1}$), we next carry out an induction for the proof of (\ref{equ_P1}). Suppose that we have
$c_i=d_i,c_{N-1-i}=-d_{N-1-i}$ for all $i\in \{0,1,\cdots,k-1\}$ where $k\leq Z-1$.
 Next, we calculate the aperiodic correlation sums at $\tau=N-k-1$ which involves the correlations of the following vectors.
\begin{equation}\label{equ2_P1}
\left [
\begin{matrix}
\mathbf{c}_1\\
\mathbf{c}_2\\
\mathbf{d}_1\\
\mathbf{d}_2
\end{matrix}
\right ]=
\left [
\begin{matrix}
1 & c_{1} & c_2 & \cdots & c_{k-1} & c_k\\
c_{N-k-1} & c_{N-k} & c_{N-k+1} & \cdots & c_{N-2} & c_{N-1}\\
1 & d_{1} & d_2 & \cdots & d_{k-1} & d_k\\
d_{N-k-1} & d_{N-k} & d_{N-k+1} & \cdots & d_{N-2} & d_{N-1}
\end{matrix}
\right ].
\end{equation}
Note that
\begin{equation}
\begin{split}
  & \rho(\mathbf{c})(N-k-1)+\rho(\mathbf{d})(N-k-1)\\
  = & \Bigl < \mathbf{c}_1,\mathbf{c}_2\Bigl >+\Bigl < \mathbf{d}_1,\mathbf{d}_2\Bigl >=0,
 \end{split}
\end{equation}
\begin{equation}
\begin{split}
  & \rho(\mathbf{c},\mathbf{d})(N-k-1)+\rho(\mathbf{d},\mathbf{c})(N-k-1)\\
  = & \Bigl < \mathbf{c}_1,\mathbf{d}_2\Bigl >+\Bigl < \mathbf{d}_1,\mathbf{c}_2\Bigl >=0.
 \end{split}
\end{equation}
By (\ref{equ2_P1}), we have
\begin{equation}\label{equ3_P1}
c^*_{N-k-1}+\sum\limits_{i=1}^{k}c_ic^*_{N-k-1+i}+d^*_{N-k-1}+\sum\limits_{i=1}^{k}d_id^*_{N-k-1+i}=0,
\end{equation}
and
\begin{equation}\label{equ4_P1}
d^*_{N-k-1}+\sum\limits_{i=1}^{k}c_id^*_{N-k-1+i}+c^*_{N-k-1}+\sum\limits_{i=1}^{k}d_ic^*_{N-k-1+i}=0.
\end{equation}
Summing up (\ref{equ3_P1}) and (\ref{equ4_P1}), we have
\begin{equation}\label{equ5_P1}
\begin{split}
 & 2(c^*_{N-k-1}+d^*_{N-k-1})+\sum\limits_{i=1}^{k}c_i\left(\underbrace{c^*_{N-k-1+i}+d^*_{N-k-1+i}}_{=0}\right)\\
 & ~~~+\sum\limits_{i=1}^{k}d_i\left(\underbrace{c^*_{N-k-1+i}+d^*_{Z-k-1+i}}_{=0}\right)=0.
\end{split}
\end{equation}
Subtracting (\ref{equ3_P1}) by (\ref{equ4_P1}), we have
\begin{equation}\label{equ6_P1}
\begin{split}
 & (c_k-d_k)c^*_{N-1}+\sum\limits_{i=1}^{k-1}\left(\underbrace{c_i-d_i}_{=0}\right)c^*_{N-k-1+i}-(c_k-d_k)d^*_{N-1}\\
 &~~~+\sum\limits_{i=1}^{k-1}\left(\underbrace{d_i-c_i}_{i=0}\right)d^*_{Z-k-1+i}=0.
\end{split}
\end{equation}
By (\ref{equ5_P1}), we have $c_{N-k-1}=-d_{N-k-1}$. By (\ref{equ6_P1}) and considering $c_{N-1}\neq d_{N-1}$\footnote{Otherwise, we have $c_{N-1}= d_{N-1}=0$ by $c_{N-1}+ d_{N-1}=0$.
This contradicts with the precondition that $(\mathbf{c},\mathbf{d})$ is a $q$-ary CZCP.}, we assert that $c_k=d_k$. Continuing this induction, the proof of (\ref{equ_P1}) follows.

Next, we assume $Z>N/2$ and $N$ is even. Setting $i=N/2-1$ into (\ref{equ_P1}), we have $c_{N/2}+d_{N/2}=0$. Then, setting $i=N/2$ into (\ref{equ_P1}),
we obtain $c_{N/2}=d_{N/2}$. This requires that $c_{N/2}=d_{N/2}=0$ which contradicts with the precondition that $(\mathbf{c},\mathbf{d})$ is a $q$-ary CZCP.
When $N$ is odd, another contradiction is reached for $c_{(N-1)/2}=d_{(N-1)/2}=0$. In either case, we assert that $Z\leq N/2$ should be met.
\end{proof}

\vspace{0.1in}
\noindent \textit{Proof of P2}:
\begin{proof}
In Section II, we have shown that if  $(\mathbf{a},\mathbf{b})$ is a GCP, $(\underline{\mathbf{b}}^*,-\underline{\mathbf{a}}^*)$ will also be a GCP which is mutually orthogonal to $(\mathbf{a},\mathbf{b})$. The first identity of (\ref{MSE_min_ACCFcond2}) is similar to the mutually orthogonal property of GCPs, although with some slight changes for its proof. Here, we only prove the second equality of (\ref{MSE_min_ACCFcond2}). For $1\leq\tau\leq Z$, we have
\begin{equation}
\begin{split}
 & \rho(\mathbf{b},\underline{\mathbf{b}}^*)(N-\tau)+\rho(\mathbf{a},-\underline{\mathbf{a}}^*)(N-\tau)\\
 = &\sum\limits_{i=0}^{\tau-1}\Bigl( b_ib_{\tau-1-i}-a_i a_{\tau-1-i}\Bigl).
\end{split}
\end{equation}
By assumption, we have $a_i=b_i,a_{N-1-i}=-b_{N-1-i},~\text{for all}~i\in(\mathcal{T}_1-1)$. Thus, the above equation reduces to
\begin{equation}\label{equ3_TypeII}
\begin{split}
 & \rho(\mathbf{b},\underline{\mathbf{b}}^*)(N-\tau)+\rho(\mathbf{a},-\underline{\mathbf{a}}^*)(N-\tau)\\
 = & \sum\limits_{i=0}^{\tau-1}a_i\Bigl( {\underbrace{b_{\tau-1-i}- a_{\tau-1-i}}_{=0}}\Bigl)=0.
\end{split}
\end{equation}
On the other hand, when $\tau=Z+1$, we have
\begin{equation}\label{equ4_TypeII}
\begin{split}
 & \rho(\mathbf{b},\underline{\mathbf{b}}^*)(N-\tau)+\rho(\mathbf{a},-\underline{\mathbf{a}}^*)(N-\tau)\\
 = &\sum\limits_{i=1}^{\lambda-1}a_i\Bigl( {\underbrace{b_{Z-i}- a_{Z-i}}_{=0}}\Bigl)+ 2a_0(\underbrace{b_{Z}-a_{Z}}_{\neq 0})\neq 0.
\end{split}
\end{equation}
This proves the second equality of (\ref{MSE_min_ACCFcond2}).
\end{proof}

\vspace{0.1in}
\noindent \textit{Proof of P3}:
\begin{proof}
For binary sequence pair $(\mathbf{a},\mathbf{b})$ over $\{-1,1\}$, by the first condition specified in (\ref{MSE_min_ACCFcond}), we have 
\begin{equation}\label{ACF_tau1}
\rho\left(\mathbf{a}\right)(N-1)+\rho\left(\mathbf{b}\right)(N-1)=a_0a_{N-1}+b_0b_{N-1}=0.
\end{equation}
Also, we require 
\begin{equation}\label{ACF_tau1_}
\rho\left(\mathbf{a}\right)(1)+\rho\left(\mathbf{b}\right)(1)=\sum\limits_{i=0}^{N-2}(a_ia_{i+1}+b_ib_{i+1})=0.
\end{equation}
Let $a_i=1-2\tilde{a}_i,b_i=1-2\tilde{b}_i$, where $1\leq i\leq N$ and $\tilde{a}_i,\tilde{b}_i\in \mathbb{Z}_2$.
Consequently, (\ref{ACF_tau1}) and (\ref{ACF_tau1_}), respectively, imply that
\begin{equation}\label{ACF_tau1_v2}
\tilde{a}_0+\tilde{a}_{N-1}+\tilde{b}_0+\tilde{b}_{N-1}= 1~(\text{mod}~2),
\end{equation}
and
\begin{equation}\label{ACF_tau1_v2_}
\begin{split}
N-1 & = \sum\limits_{i=0}^{N-2}(a_i+a_{i+1}+b_i+b_{i+1})~(\text{mod}~2)\\
    & = \tilde{a}_0+\tilde{a}_{N-1}+\tilde{b}_0+\tilde{b}_{N-1}~(\text{mod}~2).
\end{split}
\end{equation}
Combining (\ref{ACF_tau1_v2}) and (\ref{ACF_tau1_v2_}), we assert that $N$  should be an even positive integer.

In the end, (\ref{equ_P3}) of P3 can be easily obtained by an induction exploiting $\rho\left(\mathbf{a}\right)(\tau)+\rho\left(\mathbf{b}\right)(\tau)$
starting from $\tau=N-1$ to $\tau=N-Z$. This induction is similar to that in \cite{Golay61,Liu-2014} and therefore detailed proof of (\ref{equ_P3}) is omitted.
\end{proof}


\end{document}